\documentclass[a4paper,11pt]{article}
\pdfoutput=1
\usepackage{jcappub}
\usepackage{caption}
\usepackage{hyperref}
\usepackage{amsmath}
\usepackage{amstext} 
\usepackage{mathtools}
\usepackage{tablefootnote}
\usepackage{soul}
\usepackage{ulem}
\usepackage{url}
\usepackage{color}
\usepackage{verbatim}
\usepackage{float}
\usepackage{graphicx}
\usepackage{subcaption}
\usepackage[usenames,dvipsnames,svgnames,table]{xcolor}
\usepackage{makecell}
\usepackage{comment}

\title{Momentum transfer in the dark sector and lensing convergence in upcoming galaxy surveys
} 

\author[1]{Wilmar Cardona,}
\emailAdd{wilmar.cardona@unesp.br}
\affiliation[1]{ICTP South American Institute for Fundamental Research \& Instituto de F\'isica Te\'orica, Universidade Estadual Paulista, 01140-070, S\~ao Paulo, Brazil}

\author[2]{David Figueruelo}
\emailAdd{davidfiguer@usal.es}
\affiliation[2]{Departamento de F{\'i}sica Fundamental and IUFFyM, Universidad de Salamanca, E-37008 Salamanca, Spain.}

\abstract{We investigated a cosmological model that allows a momentum transfer between dark matter and dark energy. The interaction in the dark sector mainly affects the behaviour of perturbations on small scales while the background evolution matches the $w$CDM solution. As a result of the momentum transfer, these kinds of models help alleviating the $\sigma_8$ discrepancy in the standard model, but do not resolve the so-called $H_0$ tension. We confirm that this is indeed the case by computing cosmological constraints.  While our analysis tends to favour $\sigma_8$ values lower than in $\Lambda$CDM, we do not find evidence for a non-vanishing momentum transfer in the dark sector. Since upcoming galaxy surveys will deliver information on scales and red-shift relevant for testing models allowing momentum transfer in the dark sector, we also carried out forecasts using different survey configurations. We assessed the relevance of neglecting lensing convergence $\kappa$ when modelling the angular power spectrum of number counts fluctuations $C_\ell^{\rm ij}(z,z')$. We found that not including $\kappa$ in analyses leads to biased constraints ($\approx 1-5\,\sigma$) of cosmological parameters even when including information from other experiments. Incorrectly modelling $C_\ell^{\rm ij}(z,z')$ might lead to spurious detection of neutrino masses and exacerbate discrepancies in $H_0$ and $\sigma_8$.  
}
	
\begin{document}

\maketitle

\section{Introduction}
\label{section:introduction}

Finding the reason why the expansion of the Universe is speeding up remains one the biggest challenges in cosmology. Although there is compelling observational evidence from different probes such as supernovae type Ia~\cite{SupernovaCosmologyProject:1998vns,SupernovaSearchTeam:1998fmf,SDSS:2014iwm,Scolnic:2017caz}, Baryon Acoustic Oscillations (BAO)~\cite{2011MNRAS.415.2892B}, large-scale structure~\cite{SDSS:2003eyi,PhysRevLett.122.171301}, weak lensing~\cite{Hildebrandt:2016iqg}, and Cosmic Microwave Background (CMB) radiation~\cite{Aghanim:2018eyx}, our theoretical description of the phenomenon lacks in fundamental grounds. While the standard cosmological model $\Lambda$-Cold-Dark-Matter ($\Lambda$CDM) fits very well most astrophysical observations, we must bear in mind that $\Lambda$CDM is just a pretty good phenomenological description of observations with two big drawbacks. Firstly, effective Quantum Field Theory prediction for $\Lambda$ hugely disagrees from observations, giving rise to the so-called Cosmological Constant Problem~\cite{RevModPhys.61.1,Carroll:2000fy}. Secondly, we are still pretty uncertain of the nature of Cold Dark Matter (CDM) since its detection remains elusive in  laboratories~\cite{RevModPhys.90.045002,PhysRevLett.126.091101,Green:2021jrr}, only having evidence for CDM through its gravitational effects. 

The concordance model $\Lambda$CDM is successful at describing observations but lacks a fundamental theory supporting it. Therefore alternative models having a more sound theoretical ground have emerged in the literature. Matter fields in theories beyond the Standard Model of Particle Physics could fill the gap in the energy budget and provide a plausible explanation for the late-time accelerating expansion without a cosmological constant. This appealing approach is generally dubbed Dark Energy (DE)~\cite{Copeland:2006wr}, but thus far no new fields have been detected neither in the laboratory nor in astrophysical measurements. Another popular approach intends to explain the current accelerating expansion through modifications in the theory of gravity~\cite{Clifton:2011jh}. Although there are reasons to believe that General Relativity (GR) might not hold under certain conditions, GR is in very good agreement with observations and remains a key ingredient in the standard cosmological model~\cite{PhysRevLett.116.221101,Collett:2018gpf,PhysRevLett.121.231102,PhysRevLett.121.231101,PhysRevLett.125.191101,PhysRevLett.125.141104,Planck:2015bue}. 

In this work we will assume both DE and CDM exist as new, yet directly undetected fields. In the literature usually DE and CDM are allowed to interact with each other only through gravity. Here, however, we will drop this assumption and consider a possible non-vanishing interaction in the dark sector. Cosmological models where DE and CDM are coupled have been investigated in some works (see, for instance,~\cite{Amendola:1999er,Farrar:2003uw,Pourtsidou:2013nha,Boehmer:2015sha,Koivisto:2015qua,Boehmer:2015kta,Wang:2016lxa,Benetti:2019lxu}). Among all the plethora of interacting models, there is an interesting group which leave the background cosmology as in the standard model while involving a momentum transfer between two dark  fluids~\cite{Simpson:2010vh,Pourtsidou:2016ico,Linton:2021cgd,Amendola:2020ldb,Chamings:2019kcl,BeltranJimenez:2020qdu,Vagnozzi:2019kvw,Jimenez:2021ybe,Ferlito:2022mok,Kumar:2017bpv}. The preservation of the standard model at the background level is an appealing property since we know $\Lambda$CDM provides a good fit for most data sets. In this work we will focus on a cosmological model proposed recently where DE and CDM are allowed to interact via a non-vanishing Thomson-like scattering. This interaction turns out to be pretty interesting: it would act mainly on small scales and it implies a momentum transfer between DE and CDM, which could alleviate current discrepancies in cosmological parameters such as $\sigma_8$ as was shown in Refs.~\cite{Asghari:2019qld,Figueruelo:2021elm,Jimenez:2021ybe}. Non-linear scales  have been recently investigated as possible solutions for the tension in the strength of matter clustering~\cite{10.1093/mnras/stac2429}. 

Upcoming galaxy surveys are expected to provide maps of matter distribution in the Universe with enough resolution to test cosmological models on small scales. New, sophisticated data sets will require careful modelling of physical phenomena if biases in the determination of cosmological parameters want to be avoided. A general relativistic description of galaxy clustering takes into consideration that the observed galaxy fluctuation field contains additional contributions arising from the distortion in observable quantities (e.g., observed redshift and position of galaxies)~\cite{Yoo:2009au,PhysRevD.84.063505,Challinor:2011bk}. Recent investigations have assessed the relevance these new relativistic contributions will have in analyses of galaxy clustering, the conclusion being that lensing convergence should be taken into account. Otherwise relevant cosmological parameters such as neutrino masses or DE equation of state could be heavily misestimated~\cite{PhysRevD.83.123514,Duncan:2013haa,Cardona:2016qxn,Cardona:2019qaz,PhysRevD.97.023537,Euclid:2021rez}.

Galaxy clustering analyses can be carried out using either matter power spectrum $P(k,z)$ or angular matter power spectrum $C_\ell(z,z')$. While $P(k,z)$ is used in standard analyses, it has the disadvantage of not being directly observable~\cite{PhysRevD.84.063505}. An analysis using the power spectrum in harmonic space $C_\ell(z,z')$ has at least three main advantages over an analysis employing $P(k,z)$: i) it is an observable~\cite{PhysRevD.84.063505}; ii) it is frame independent~\cite{Francfort:2019ynz}; iii) it is relatively easy to take into consideration relativistic effects~\cite{DiDio:2013bqa}.  In this work we perform forecasts for an EUCLID-like galaxy survey by utilising the angular matter power spectrum. We investigate how relevant is the effect of lensing convergence when analysing galaxy number counts in the context of a dynamical DE model where DE and DM are allowed to interact with each other via a Thomson-like scattering. We determine  whether or not neglecting lensing convergence, while considering a small scales DM-DE interaction, hinders accurate estimation of cosmological parameters. The paper is organised as follows. In Section~\ref{section:theory} we explain the interacting model under consideration in this work, briefly discuss its phenomenology, and recall the general expressions for a general relativistic description of galaxy clustering. In Section~\ref{section:methodology} we provide details about data sets as well as the Markov Chain Monte Carlo (MCMC) technique we use to compute cosmological constraints and forecasts. Sections~\ref{section:results} and~\ref{section:discussion} are dedicated to our results and their relation to relevant literature, respectively. Finally in Section~\ref{section:conclusions} we conclude. 

\section{Theoretical framework}
\label{section:theory}

\subsection{Covariantised Thomson-like dark scattering}
\label{subsection:model}

In this section, we introduce the general framework of the interacting model under consideration. For more details we refer the reader to Refs.~\cite{Asghari:2019qld,Figueruelo:2021elm} where a full derivation is performed.

We assume the standard Friedmann–Lemaître–Robertson–Walker metric (FLRW) and that the matter/energy components of the Universe can be described by perfect fluids, with all the non-interacting fluids following the standard conservation law of its stress-energy tensor $\nabla_\mu T_{(\rm{i})}^{\mu \nu}=0$, with $\rm{i}$ representing baryons, photons, etc.
The dark components of the Universe are described by the interacting model, first derived in Ref.~\cite{Asghari:2019qld}, where an interaction driven by the relative motion of dark energy (DE) and dark matter (DM) is introduced. This kind of interaction can be understood as a covariantisation of a dark Thomson scattering similar to the baryon-photon fluid prior to recombination [see Eqs.~\eqref{eq:deltaDM}-\eqref{eq:thetaDE}].
Interestingly, this kind of interaction does not modify neither the background nor the continuity equations of the coupled fluids. The interaction introduces a new term in the Euler equations of each coupled fluid proportional to their relative velocity. We can formalise this interaction for the DE-DM coupling by the following non-conservation equations 
\begin{eqnarray}
\nabla_\mu T^{\mu\nu}_c &=&\Bar{\alpha}(u^\nu_c-u^\nu_d)\,,\\
\nabla_\mu T^{\mu\nu}_d&=&
-\Bar{\alpha}(u^\nu_c-u^\nu_d)\,,
\label{eq:Defintalpha}
\end{eqnarray}
where $\Bar{\alpha}$ describes the strength of the interaction, that we assume constant for simplicity; $T^{\mu\nu}_c$ and $T^{\mu\nu}_d$ are the energy-momentum tensors for DM and DE, respectively; $u^\nu_c$ and $u^\nu_d$ denote $4$ velocities for DM and DE, respectively. For convenience, we will use a dimensionless coupling constant defined as 
\begin{equation}
    \alpha = \frac{\Bar{\alpha}}{\rho_{\rm crit} H_0} ,
\end{equation}
with $\rho_{\rm crit}=\frac{3H_0^2}{8\pi G}$ the critical density and $H_0$ the Hubble parameter today. This normalisation will lead to ranges of the coupling of order one $\alpha\sim\mathcal{O}(1)$.

The coupling clearly has no impact on the background cosmology as both fluids share the same rest frame on sufficiently large scales: $u^\nu_c \simeq u^\nu_d$. However, when considering linear perturbations the DE-DM coupling plays a relevant role. Assuming the linearly perturbed FLRW metric in the Newtonian gauge\footnote{Since in this work we neglect DE anisotropic stress, the gravitational potentials satisfy $\Phi=\Psi$.}, the linear perturbations equations for the density contrast $\delta\equiv\delta\rho/\rho$ and the Fourier space velocity $\theta\equiv i\Vec{k}\cdot\Vec{v}$  are
\begin{eqnarray}
\label{eq:deltaDM}
\delta_{c}' &=& 
-\theta_{c}+3\Phi'\,,\\
\delta_{d}'&=&-3 \mathcal{H}( c_s^2-w) \delta_{d} +3(1+w)\Phi'  -(1+w)\left(1+9 \mathcal{H}^2
\frac{c_s^2-w}{k^2}\right)
\theta_d\;, \\
\label{eq:thetaDM}
\theta_c'&=&-\mathcal{H} \theta_c + k^2 \Phi + \Gamma(\theta_d-\theta_c)\;, \\
\label{eq:thetaDE}
\theta_d' &=&(3c_s^2-1) \mathcal{H}\theta_d+k^2\Phi +\frac{k^2 c_s^2}{1+w}\delta_d-\Gamma R_{cd}(\theta_d-\theta_c)\,,
\end{eqnarray}
where $w$ is the constant DE equation of state, $c_s^2$ is the squared sound speed of DE defined in the rest frame which we fix to $c_s^2=1$, $\mathcal{H}$ is the conformal Hubble function and 
\begin{eqnarray}
\Gamma&\equiv& \alpha \frac{a}{\rho_c} \;, \\ \label{eq:Scoupling}
R_{cd} &\equiv&
\frac{\rho_c}{(1+w)\rho_d}\,, \label{eq:Rcoupling}
\end{eqnarray}
represent the dark sector interaction rate and the dark sector energy ratio, respectively.

It is worth saying that this interaction becomes efficient when $\Gamma>\mathcal{H}$. Since we are considering a constant coupling $\alpha$ we have $\Gamma \propto a^4$ and $\Gamma R \propto a^{4+w}$, then we should see the effect of the interaction in the late-time Universe when these terms would be large enough to dominate the evolution of the perturbations. Moreover, the interaction is only relevant on small scales provided it needs a non-zero relative velocity between both interacting components to be efficient, while on large scales both fluids have the same rest frame, hence the coupling term of equations~\eqref{eq:thetaDM}-\eqref{eq:thetaDE} vanishes. 

\begin{figure}[http]
	\centering
	\includegraphics[scale=0.1675]{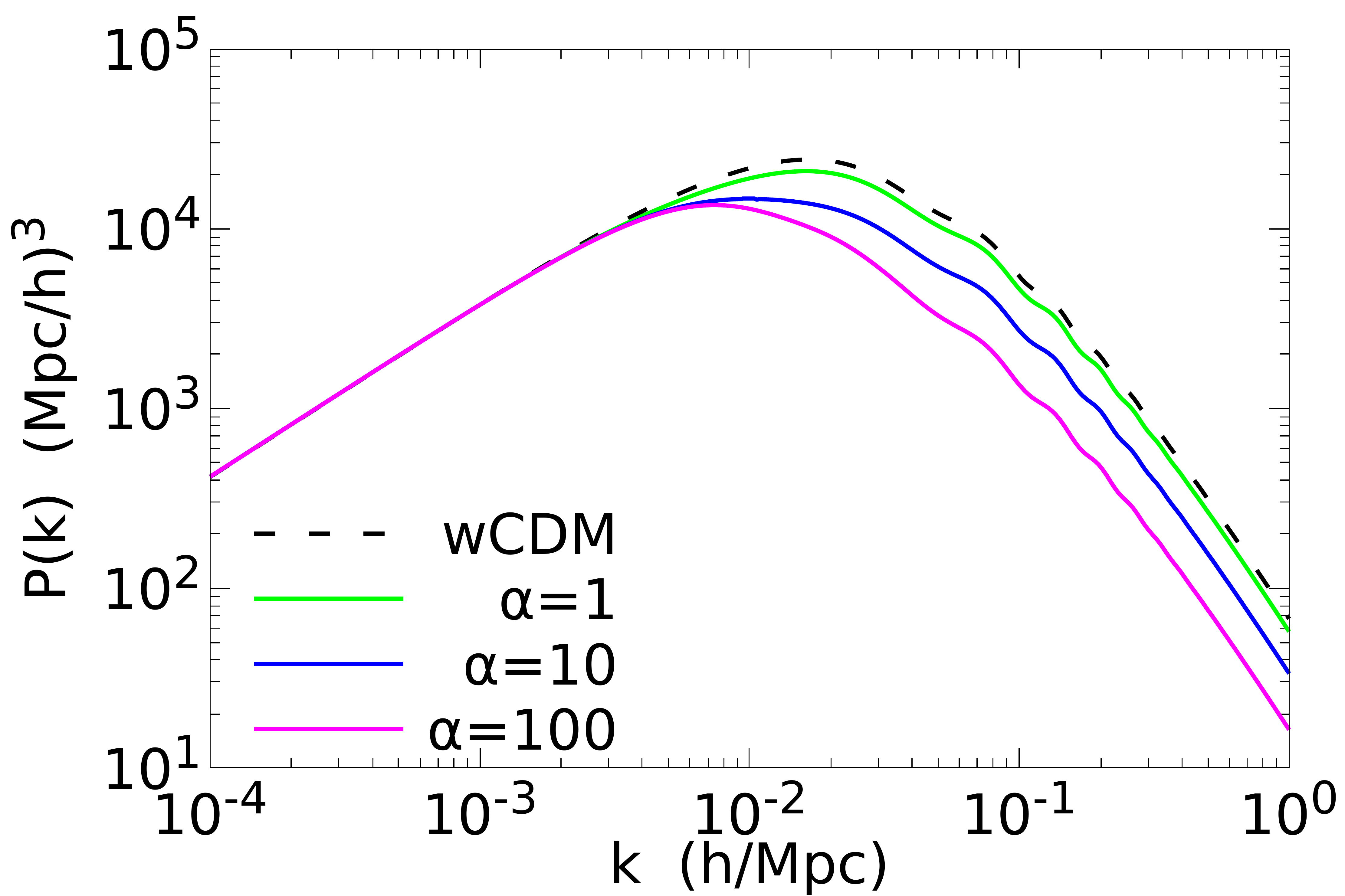}
	\includegraphics[scale=0.16]{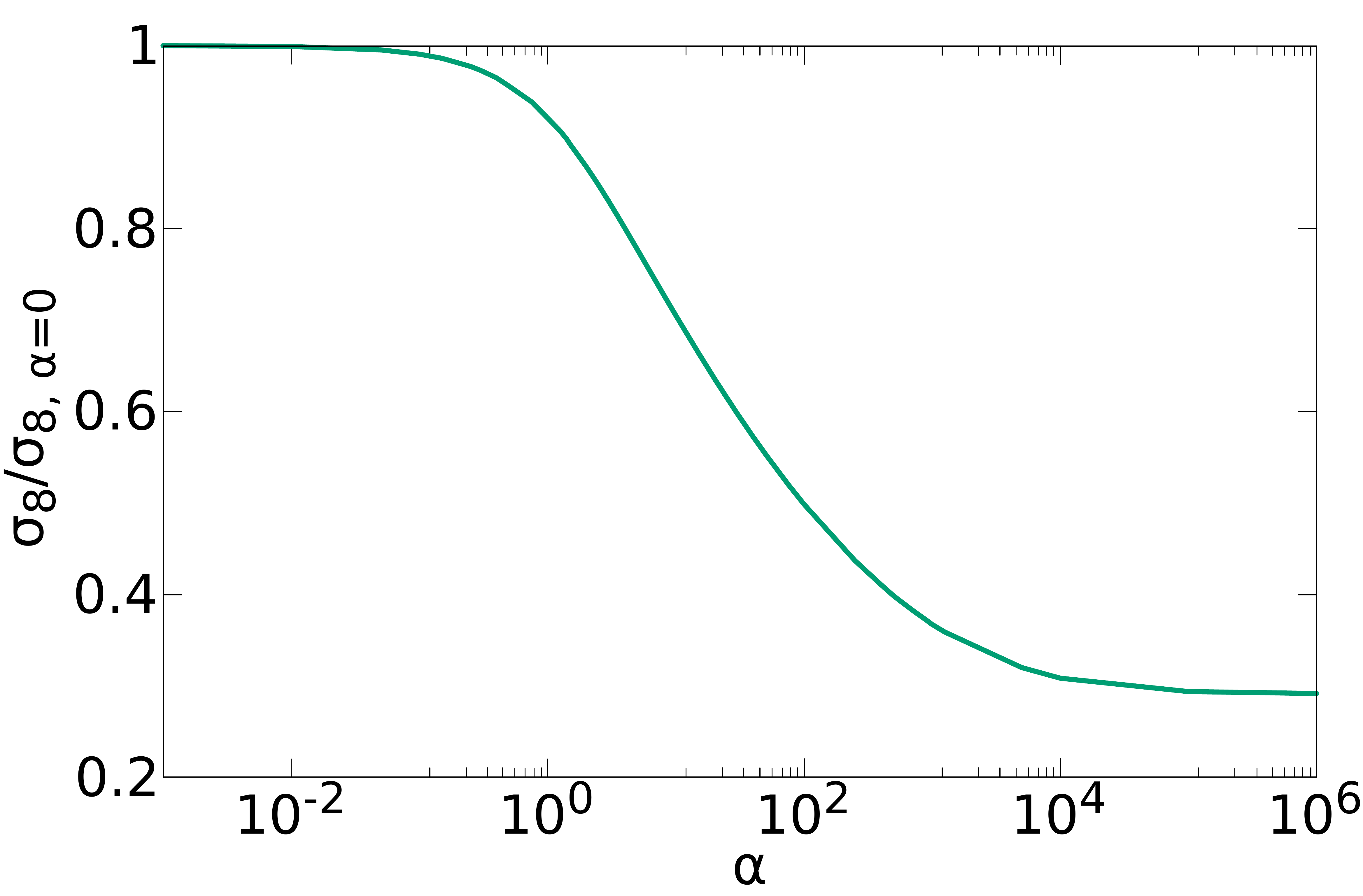}
	\caption{In the left plot, we show the matter power spectrum for the reference $w$CDM model (black) and for the elastic model for values of the coupling parameter $\alpha=1$ (green), $\alpha=10$ (blue) and $\alpha=100$ (pink). In the right plot, we show the ratio between $\sigma_8$ for several values of $\alpha$ and its non-interacting value $\sigma_{8,\alpha=0}$.}
	\label{Fig:PK}
\end{figure}

\begin{figure}[http]
	\centering
	\includegraphics[scale=0.16]{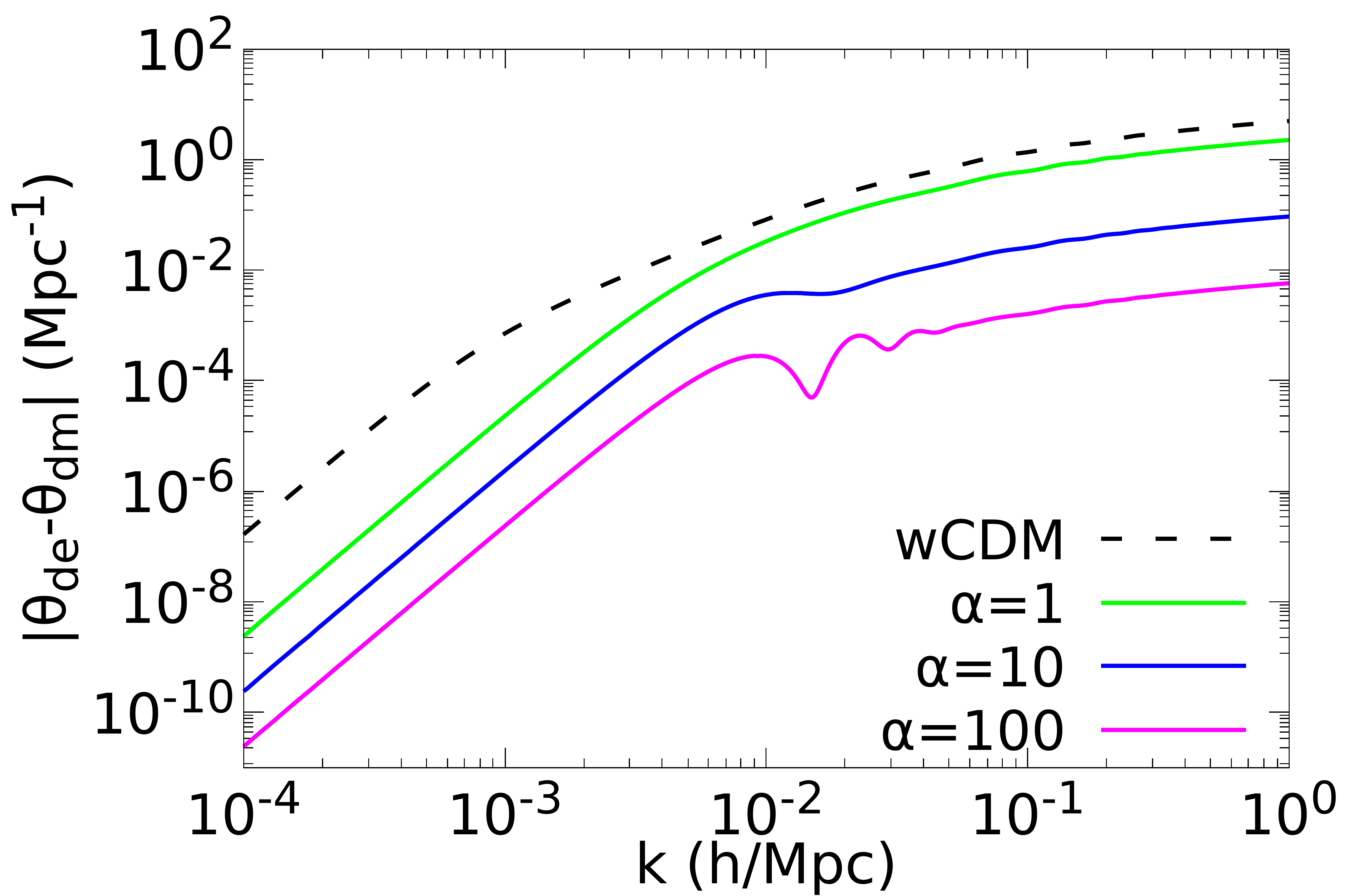}
	\includegraphics[scale=0.16]{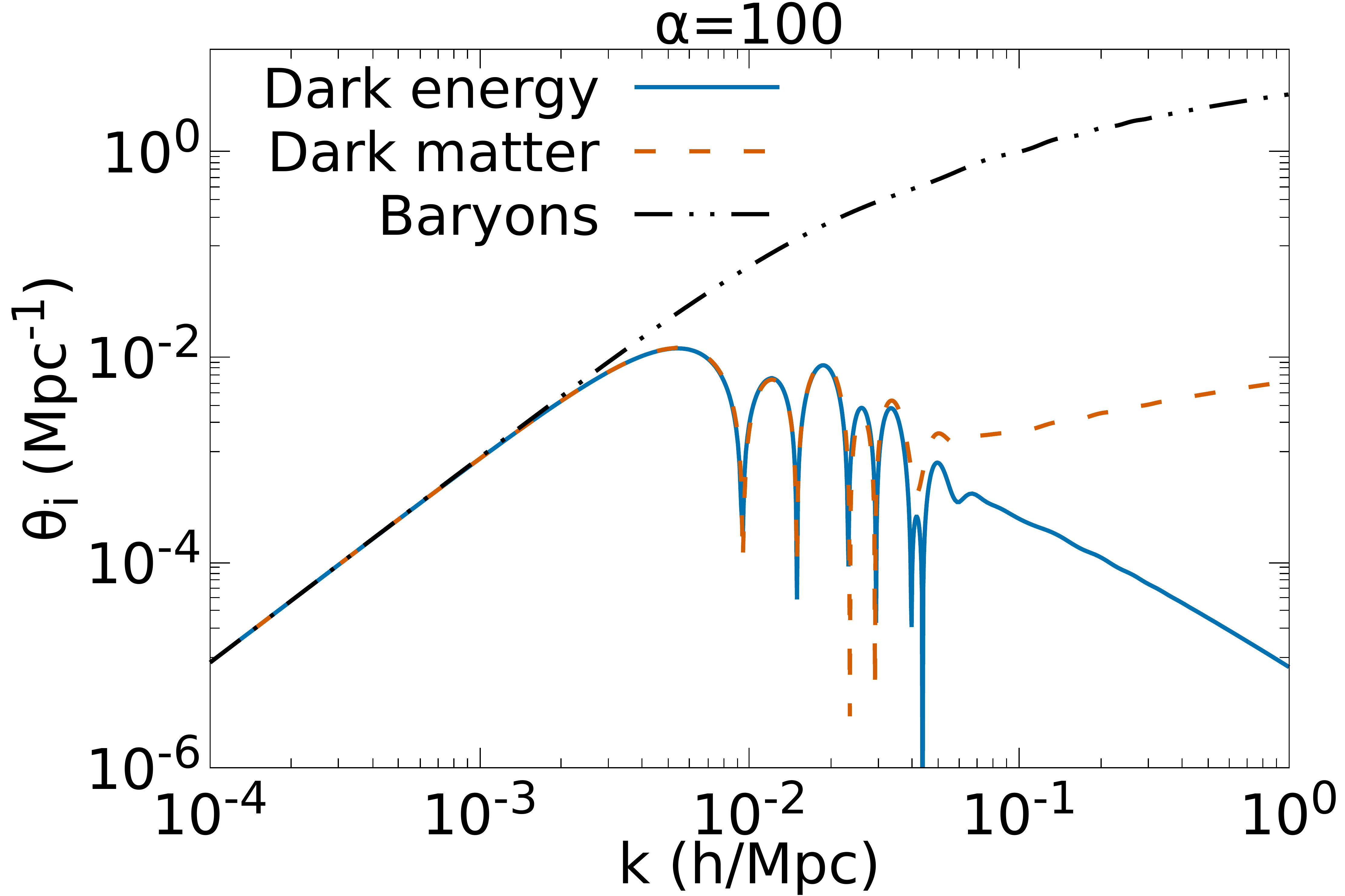}
    \caption{In the left plot we show the relative velocity between dark energy and dark matter for the reference $w$CDM model and for the elastic model with values of the coupling parameter $\alpha=1,10,100$. In the right plot, we show the velocity of dark energy, dark matter and baryons for the elastic model with $\alpha=100$. Both plots illustrate the Dark Acoustic Oscillations due to the momentum transfer in the dark sector.}
	\label{Fig:DAO}
\end{figure}

Although the model has a rich phenomenology (see Refs.~\cite{Asghari:2019qld,Figueruelo:2021elm}), here we will only focus on the most relevant effects that we study via an implementation in the Boltzmann code~\texttt{CLASS}~\cite{2011JCAP...07..034B}. For illustrative purposes about the effects of this model, in the following plots (Figures~\ref{Fig:PK},~\ref{Fig:DAO} and~\ref{fig:cl}) we use as cosmological parameters $H_0=67.37\; km/s/Mpc$, $\Omega_{\rm b} h^2 = 0.02246$, $\Omega_{\rm c} h^2 = 0.119$, $n_{\rm s} = 0.9679$, $A_{\rm s}=2.099\;10^{-9}$, $\tau=0.054$ and $w=-0.978$ whereas we allow to have one massive neutrino with $m_\nu=0.31$ while the other two are massless.
Firstly, the model predicts a suppression on the matter power spectrum $P(k)$ on small scales. When the interaction is efficient the evolution of DM perturbations departs from the standard case as DE pressure acts on DM slowing down the growth of DM density perturbations due to gravitational collapse on small scales.\footnote{This is a similar process to baryon-photon Thomson scattering before recombination, when radiation pressure of photons prevents the clustering of the  pressureless baryons. In our scenario DE accounts for the pressure while DM plays the role of a pressureless fluid which is prevented from clustering. There is a crucial difference, though: while Thomson scattering took place when there were no gravitationally bound structures, the elastic interaction happens recently when gravity has already created a lumpy universe.} Then, DM structures stop growing and this is imprinted in the matter power spectrum as a suppression at small scales $k\sim10^{-2}-10^{0}\, \mathrm{h/Mpc}$ (for realistic values of $\alpha$ allowed by current constraints of galaxy surveys \cite{DES:2021wwk}) while larger scales remain as in the standard case, as we can see in the left panel of Figure~\ref{Fig:PK}. As a consequence of the freezing of DM perturbations, less clustering is expected. This has a direct impact on the parameter $\sigma_8$, which captures the clustering at $8\rm{h}^{-1}\, \mathrm{Mpc}$ scales, resulting in a lower value depending on the model parameter $\alpha$ as seen in the right panel of Figure~\ref{Fig:PK}. As a side effect due to the freezing of dark matter perturbations, the gravitational potential acquires a time dependence that will contribute to a late time Integrated Sachs–Wolfe effect. However, for $\alpha \sim 1$ these changes would be tiny and only relevant for very large scales (low $\ell$ modes) where cosmic variance dominates the error budget, hence not very informative for cosmological constraints analyses.

Secondly, an interesting feature of the elastic model is the prediction of a shift in the turnover of the matter power spectrum. The peak is usually determined by the horizon at matter-radiation equality. Since the elastic model does not alter the background dynamics with respect to the standard evolution, it is naively expected that the turnover occurs at the same equality scale. Nevertheless, a non-vanishing interaction in the dark sector changes the picture. Once the momentum exchange becomes relevant DM and DE behave as a single fluid and the growth of structures freezes. As a result there is a shift in the turnover of the matter power spectrum due to modes entering the horizon after the interaction is switched on (see Fig.~\ref{Fig:PK}). 

Following the similarity between this interaction and the pre-recombination Thomson scattering between photons and baryons which provokes the Baryons Acoustic Oscillations (BAO), the elastic coupling leads to the emergence of the Dark Acoustic Oscillations (DAO), as we can see in the relative velocity between the coupled components shown in the left panel of Figure~\ref{Fig:DAO}. However, it is worth saying the relevant scales are different for DAO and BAO: while BAO is a early Universe process, DAO is a late time effect; also  DAO appear for scales larger than BAO around $k\sim 10^{-2}\; \rm{h}\,\rm{Mpc}^{-1}$.
In the right panel of Figure~\ref{Fig:DAO}, the oscillation regime corresponds to the scales where the interaction is strongly efficient, namely, where both fluids are strongly coupled. For smaller scales the interaction cannot overcome the gravitational collapse despite the fact it is still able to drag DM.

The presence of a non-vanishing momentum exchange in the dark sector was investigated in Ref.~\cite{Asghari:2019qld}. Authors found $>3\sigma$ evidence for a non-vanishing elastic interaction between DM and DE.  Nevertheless, this result was sharpened up in Ref.~\cite{Figueruelo:2021elm} showing how the Sunyaev–Zeldovich (\texttt{SZ}) cluster count likelihood\footnote{This likelihood acts as a Gaussian prior on the combination of parameters $S_8$ defined as~$S_{8,{\rm SZ}}\equiv\sigma_8\left(\Omega_m/0.27\right)^{0.3}=0.782\pm0.010\,$, implemented in~\texttt{MontePython}~\cite{Audren:2012wb,Brinckmann:2018cvx} under the name of \texttt{Planck\_{}SZ}, from Table 2 of Ref.~\cite{Ade:2013lmv} obtained from the combination of~\texttt{Planck2013+BAO+BBN} data  with a fixed mass biased $1-b=0.8$. } has a key role in the claimed detection. The \texttt{SZ} likelihood relies on small scale simulations using $\Lambda$CDM as a fiducial model which might not lead to fully consistent results when other models are investigated. In this work we adopt a conservative position and choose not to include the \texttt{SZ} likelihood when computing cosmological constraints for the elastic model. The effect of the~\texttt{SZ} likelihood is worth of attention as similar results were found when considering other interactions where a momentum exchange takes place (see  Refs.~\cite{Pourtsidou:2016ico,Jimenez:2020ysu,Linton:2021cgd}). In Section~\ref{section:results} we will compute cosmological constraints for the interacting model under consideration using the most up-to-date data sets as well as forecasts for an Euclid-like galaxy survey. 

\subsection{Galaxy number counts}
\label{subsection:galaxy-nc}

\begin{figure}[http]
	\centering
	\includegraphics[scale=0.75]{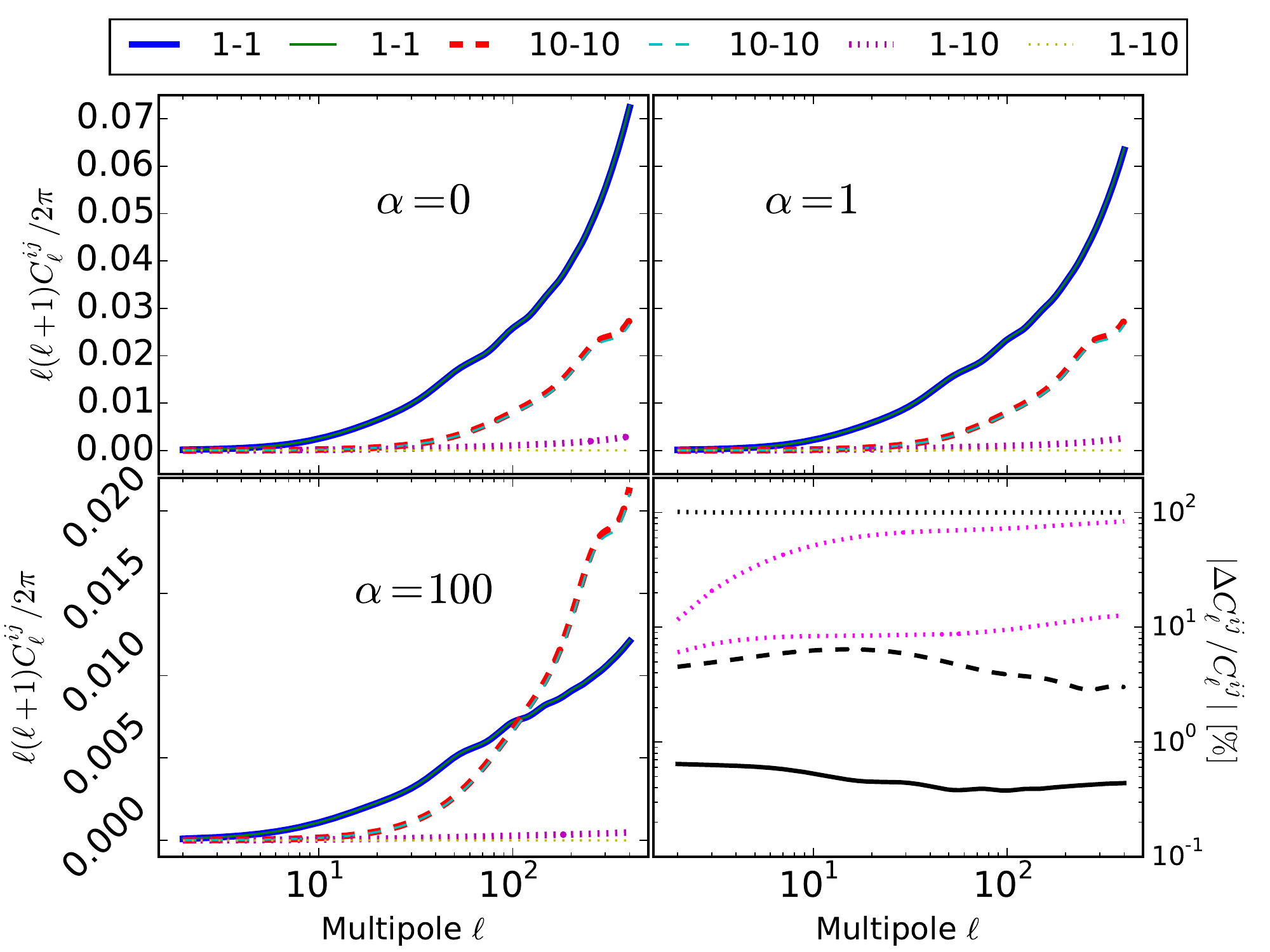}
    \caption{Galaxy number counts power spectrum (upper, right panel and left panels). The indices for the correlated redshift bins are shown in the legend. Thick and thin lines correspond to computations consistently including lensing convergence and neglecting it, respectively. We plot the observables for different values of the parameter $\alpha$ governing the DM-DE interaction ($\alpha=0$ corresponds to $w$CDM). Angular power spectra were computed using high precision parameters of Table~\ref{tab:precision} and using the~\texttt{HALOFIT} option of~\texttt{CLASS} code for the top-hat configuration. Lower, right panel: black lines shows the percentage difference when lensing convergence is neglected for the case $\alpha=1$ in the upper, right panel; magenta, dotted lines indicate the relative percentage difference between cross-correlations including lensing convergence for $\alpha=100$ (upper) and $\alpha=1$ (lower) with respect to a model with vanishing interaction.}
	\label{fig:cl}
\end{figure}

The power spectrum in the harmonic space $C_\ell(z,z')$ provides a powerful direct observable for galaxy surveys with the advantage that it is frame independent and relativistic corrections can be easily included. The observed galaxy fluctuation field has additional contributions arising from the distortion of the observed redshift $z$ and position of galaxies in the sky $\vec{n}$.  Schematically, for a mean density per redshift and per steradian $\bar{n}(z)$, the number counts are given by~\cite{Montanari:2015rga,PhysRevD.84.063505,Challinor:2011bk} 
\begin{equation}
     n(\vec{n},z)=\bar{n}(z)(1+\Delta(\vec{n},z))\; ,
\end{equation}
where the fluctuations can be written as
\begin{equation}
\Delta(\vec{n},z)=\Delta^{\rm{D}}(\vec{n},z)+\Delta^{\rm{RSD}}(\vec{n},z)+\Delta^{\rm{L}}(\vec{n},z)+\Delta^{\rm{V}}(\vec{n},z)+\Delta^{\rm{P}}(\vec{n},z) \; . \end{equation}
Here the density term is $\Delta^{\rm{D}}(\vec{n},z)=b D$ with $b$ the bias and $D$ the growth function;\footnote{As it is explained in Refs.~\cite{Asghari:2019qld,Figueruelo:2021elm}, the DM-DE coupling considered here breaks the scale independence of the growth function of the $\Lambda$CDM model. This consequence can be even inferred from the shape of the power spectrum, as this interaction leads to a scale dependent suppression of it.} the Redshift-Space Distortions (RSD) term is  $\Delta^{\rm{RSD}}(\vec{n},z)=\mathcal{H}^{-1}\partial^2_r V$ with $\mathcal{H}$ the conformal Hubble function and $V$ the velocity potential for the peculiar velocity in the longitudinal gauge, $\Delta^{\rm{L}}(\vec{n},z)=-(2-5s)\kappa$ with $s$ the magnification bias and $\kappa$ the lensing convergence defined in~\cite{PhysRevD.83.123514} as 
\begin{equation}
\kappa=\frac{3\Omega_{\rm m}H_0^2}{2}\int_0^{\chi(z)} \frac{\chi(\chi(z)-\chi)}{\chi(z)}\delta(\vec{n}\chi,\chi) d\chi,
\end{equation}
with $\chi(z)$ the comoving distance. The main contributions to the spectrum are the density fluctuations, redshift-space distortions and the lensing contributions. The density contribution usually dominates over all scales followed by the redshift-space distortions and the lensing term. The velocity term $\Delta^{\rm{V}}(\vec{n},z)$ as well as the potential term $\Delta^{\rm{P}}(\vec{n},z)$ remain subdominant for the scales considered and hence we will not take them into account when computing galaxy number counts angular power spectrum $C_\ell(z,z')$. For more details, we refer to the Refs.~\cite{Montanari:2015rga,PhysRevD.84.063505,Challinor:2011bk} where a fully analysis is done which we have followed.

The effect of the dark sector momentum exchange can be also seen in the galaxy number counts power spectrum. Figure~\ref{fig:cl} shows a set of auto-correlations and cross-correlations for different values of the coupling parameter $\alpha$. Since in this work we focus on the relevance of lensing convergence $\kappa$, we also show the number counts angular power spectrum neglecting $\kappa$. As shown in Fig.~\ref{fig:cl}, the DM-DE interaction provokes a power suppression mainly affecting auto-correlations in the lower redshift bins. Auto-correlations in the upper redshift bins are only diminished for very large values of the coupling parameter $\alpha$. This behaviour is expected as the interaction becomes only efficient for the late Universe. With regard to cross-correlations we can see that even though on small scales they are subdominant, differences with respect to the $w$CDM can reach $10\%$ for $\alpha=1$ and $90\%$ for $\alpha=100$ (see dotted, magenta lines in the lower, right panel of Fig.~\ref{fig:cl}). We can also note that while neglecting lensing convergence in auto-correlations induces a $\sim 1\%$ error in the computation of spectra, for cross-correlations the effect of lensing convergence is dominant. Neglecting lensing convergence leads to bad modelling of number counts angular power spectra. 

Let us explain the procedure of  modeling the angular power spectrum of number counts fluctuations. We assume the galaxy survey is divided into a certain number of redshift bins so that we can compute auto- and cross-correlations of number counts fluctuations. We model angular power spectrum of number counts fluctuations $C_l^{\rm{A,ij}}$ as the sum of three terms

\begin{equation}
     C_\ell^{\rm{A,ij}}= C_\ell^{\rm{ij}}+E_\ell^{\rm{ij}}+\mathcal{N}\delta^{\rm{ij}}.
     \label{eq:ClAij}
\end{equation}
Here $\rm{i}$ and $\rm{j}$ denote redshift bins, A indicates whether we are dealing with the observed or the theoretical power spectrum ($\rm{A=obs, th}$), and $\mathcal{N}$ is a shot-noise contribution due to the discreteness of our sample. The term $C_\ell^{\rm{ij}}$ represents the angular power spectrum of number counts fluctuations computed by the Boltzmann solver~\texttt{CLASS}~\cite{2011JCAP...07..034B,DiDio:2013bqa}. This code also has a few options to compute non-linear corrections. In this work we chose to do it  through the fitting function~\texttt{HALOFIT}; in Eq.~\eqref{eq:ClAij} $E_\ell^{\rm{ij}}$ corresponds to the error we make when taking in non-linear scales and we compute it as
\begin{equation}
   E_\ell^{\rm{ij}}=\mid{C_\ell^{\rm{ij}}{}^{\textit{,\rm{HALOFIT ON}}}-C_\ell^{\rm{ij}}{}^{\textit{,\rm{HALOFIT OFF}}}}\mid.
\end{equation}
In this forecast, we compute the observed angular power spectrum $C_\ell^{\rm{obs,ij}}$ and the non-linear error $E_\ell^{\rm{ij}}$ only once for a fiducial cosmology and take into consideration all the relevant contributions (i.e., lensing convergence, density perturbations and redshift space distortions)~\cite{PhysRevD.84.063505,Challinor:2011bk}. Since in this work we are interested in the relevance of lensing convergence in analyses of upcoming galaxy surveys, when carrying out the forecast we compute the theoretical angular power spectrum $C_\ell^{\rm{th,ij}}$ in two situations: i) a consistent computation where $C_\ell^{\rm{th,ij}}$ includes lensing convergence; ii) an approximate computation where lensing convergence is neglected in $C_\ell^{\rm{th,ij}}$. The observed angular power spectrum $C_\ell^{\rm{obs,ij}}$ always takes into account lensing convergence.

Computing angular power spectra of number counts fluctuations requires the implementation of the interacting model under consideration in a Boltzmann solver as well as providing specifications for the galaxy survey under consideration. In Subsection~\ref{subsection:model}, we explained the foundations of the interacting model as well as the main phenomenological consequences obtained with the modified version of the code~\texttt{CLASS}. Below we provide details of an Euclid-like galaxy survey and explain our methodology for the forecast.

\section{Methodology}
\label{section:methodology}

\subsection{Cosmological constraints using MCMC}
\label{subsection:mcmc}

We computed cosmological constraints by using recent, publicly available data sets. In order to constrain the background we included BAO measurements~(\texttt{BAO}) from Refs.~\cite{Alam:2016hwk,2011MNRAS.416.3017B,Ross:2014qpa}. Our analysis also took into consideration data sets constraining linear order perturbations. We took in CMB lensing~(\texttt{lensing}) as well as temperature and polarisation anisotropies of the CMB~(\texttt{TTTEEE}) measured by the Planck Collaboration~\cite{Aghanim:2018eyx}. 

The observables of the cosmological model (e.g., CMB angular power spectrum, matter power spectrum) were computed by using our implementation of the interacting model under consideration in the Boltzmann solver~\texttt{CLASS}. For a given set of cosmological parameters the code computes relevant background quantities and solves the set of differential equations governing the evolution of linear perturbations. Having a solution for the perturbations of each fluid, the code also computes the statistical properties we are interested in. Comparison of theoretical predictions against measurements is not an easy task because the analysis must take into account a number of nuisance parameters. In this case, an analytical treatment for extracting the statistical information and finding a best fitting model becomes hard. The usual approach in cosmology to overcome this problem and carry out the statistical analysis is via a MCMC technique~\cite{Lewis:2002ah,2017ARA&A..55..213S}. The parameter space (i.e., nuisance and cosmological parameters) is sampled with the help of the code~\texttt{Monte Python}~\cite{Audren:2012wb,Brinckmann:2018cvx} choosing the default Metropolis-Hastings algorithm. The code~\texttt{CLASS} is integrated into~\texttt{Monte Python} so that theoretical predictions are computed and compared to observations through likelihood functions several times (e.g., $\sim 10^6$ iterations). The analysis is carried out until the Gelman-Rubin statistic $R$ shows convergence \cite{10.1214/ss/1177011136}, considered  when all parameters fulfill the criterion  $R-1 \lesssim 0.01$.

In order to compute cosmological constraints we consider a cosmological model having the following varying parameters: baryon density today $\Omega_b h^2$; cold dark matter density today $\Omega_c h^2$; $100\times$ angular size of sound horizon at redshift $z_\star$ (redshift for which the optical depth equals unity) $100\theta_\star$; log power of the primordial curvature perturbations $\ln 10^{10}A_{\rm s}$; scalar spectrum power-law index $n_{\rm s}$; Thomson scattering optical depth due to reionisation $\tau$; the sum of neutrino masses  $\sum m_{\nu}(\rm{eV})$; dark energy equation of state $w$; dimensionless coupling parameter of the model $\alpha$. In our MCMC analysis we use the same prior range as specified in Table 1 of Ref.~\cite{Planck:2013pxb}, except for $\alpha$ and $w$ for which we use a flat prior range $[0,100]$ and $(-1,-0.3]$, respectively.\footnote{The condition on the equation of state $w$ avoids instabilities in the perturbations equations for a non-vanishing positive value of $\alpha$ appearing when $w<-1$, as explained in Appendix B of  Ref.~\cite{Figueruelo:2021elm}.}

\subsection{Galaxy survey specifications}
\label{subsection:survey}

We will perform the forecast for the future experiment EUCLID, a forthcoming mission of the European Space Agency. EUCLID will be placed in the L2 Sun-Earth Lagrangian point and will scan the sky during six years covering 15000~deg$^2$ for a redshift range extended up to~$z\sim2$. The mission will focus on weak lensing, BAO as well as RSD and will be equipped with two instruments, namely, the near-infrared spectrometer and photometer (NISP) and the visible imager (VIS). With those instruments, EUCLID will map the matter distribution and improve the knowledge on the evolution of the late universe and the nature of dark energy. Here, we will follow the EUCLID survey specifications in Refs.~\cite{Laureijs:2011gra,Amendola:2016saw}. 

Since we are dealing mainly with linear scales, it is plausible to assume a galaxy bias prescription which is scale-independent
\begin{equation}
b(z) = b_0 \sqrt{1+z},
\label{Eq:galaxy-bias}
\end{equation}
where the parameter $b_0$ is a constant. In the MCMC forecasts that we present in Section~\ref{section:results} we will marginalise over $b_0$.
For the magnification bias we use the following prescription
\begin{equation}
    s(z) = 0.1194+0.2122z-0.0671z^2+0.1031z^3,
\label{Eq:magnification-bias}
\end{equation}
where all the coefficients were obtained in Ref.~\cite{Montanari:2015rga}. Recently, while we were preparing this manuscript, the Euclid Collaboration extracted both galaxy bias $b(z)$ and magnification bias $s(z)$ from the Flagship simulation~\cite{Euclid:2021rez}.  

We assume that the number of galaxies per redshift and steradian follows the distribution
\begin{equation}
    \frac{dN}{dz d\Omega}=3.5\times10^8 z^2 \exp\left[{-\left(\frac{z}{z_0}\right)^\frac{3}{2}}\right],
\end{equation}
where the parameter $z_0$ is defined as  $z_0=\frac{z_{\mathrm{mean}}}{1.412}$ and the mean redshift for the EUCLID survey is taken to be $z_{\mathrm{mean}}=0.9$. Although the survey will cover up to redshift $z\sim2$, we restrict the  redshift interval to $z\in[0.1,2]$ for numerical convenience. The galaxy density is set to $d=30$ arcmin$^{-2}$ and the fraction of the sky covered is $f_{\mathrm{sky}}=0.364$.  Finally, we model the shot-noise contribution [see Eq.~\eqref{eq:ClAij}] to the spectrum as
\begin{equation}
    \mathcal{N}=N_{\mathrm{bins}}\frac{1}{3600\, d\, (\frac{180}{\pi})^2}.
\end{equation}
In our forecast we will take into consideration two configurations for the number of bins: i) $N_{\rm{bins}} = 5$ and ii) $N_{\rm{bins}} = 10$.

\subsection{Forecast using MCMC}
\label{subsection:forecast-methodology}

We carry out the forecast by following a MCMC approach. In Subsection~\ref{subsection:mcmc} we explained the procedure when computing cosmological constraints. Here, however, the analysis has a few differences. Firstly, instead of using real data for the fluctuations in number counts, we assume a set of fiducial cosmological parameters (see Table~\ref{Table:fiducial-model}) and compute the observed angular power spectra [see discussion surrounding Eq.~\eqref{eq:ClAij}]; in~\texttt{CLASS} we set the galaxy survey specifications of Subsection~\ref{subsection:survey} and use the high precision parameters in Table~\ref{tab:precision}. Secondly, we include number count fluctuations in the analysis through the likelihood function of Refs.~\cite{2013JCAP...01..026A,Cardona:2016qxn}. Here we are interested in estimating the relevance of lensing convergence so that we model number count fluctuations in two ways: i) consistently including lensing convergence; ii) neglecting lensing convergence. We compute the relative~$\chi^2$ to the observed number count fluctuations~$C_\ell^{\rm{obs,ij}}$ (which are always calculated including lensing convergence) by
\begin{table}[http]
\centering
\begin{tabular}{@{}cc}
\hline
Parameter & Value \\
\hline
$H_0\left(\frac{\text{km}}{\text{s}\cdot\text{Mpc}}\right)$ & $ 67.38 $ \\
$\Omega_b h^2$ & $ 0.02247 $ \\
$\Omega_c h^2$ & $ 0.1193 $ \\
$\tau$ & $ 0.0543 $ \\
$n_s$ & $ 0.9679 $ \\
$\ln10^{10}A_s$ & $ 3.044 $ \\
$b_0$ & $ 1 $\\
$ \sum m_\nu $\,(eV) & $ 0.031 $\\
$w$ & $-0.98$ \\
$ \alpha $ & $ 0.0723 $ \\
\hline
\end{tabular}
\caption{Parameters defining the fiducial model in our forecasts.}
\label{Table:fiducial-model}
\end{table}

\begin{table}[http]
\centering
\begin{tabular}{||c||c||c|c|c|c||}
\cline{2-6}
 \multicolumn{1}{c|}{}& $C_\ell^{\mathrm{obs,ij}}$ &  \multicolumn{2}{c|}{$C_\ell^{\mathrm{th,ij}}$ Top-hat} & \multicolumn{2}{c||}{$C_\ell^{\mathrm{th,ij}}$ Gaussian} \\\hline
Parameter & \footnotesize{All configurations} & \footnotesize{5 bins}& \footnotesize{10 bins}& \footnotesize{5 bins}& \footnotesize{10 bins}       \\\hline\hline
\footnotesize{\verb|l_switch_limber_for_nc_local_over_z|} & 20000 & 20000 &  20000 & 20000 & 20000 \\\hline
\footnotesize{\verb|l_switch_limber_for_nc_los_over_z|}   & 1000  & 1000  &  1000  & 1000  & 1000 \\\hline
\footnotesize{\verb|selection_sampling_bessel|}           & 3     & 1.2   &  1.2   & 1.2   & 1.2 \\\hline
\footnotesize{\verb|q_linstep|}                           & 0.3   & 2.5   &  1.65  & 40    & 10 \\\hline
\footnotesize{\verb|k_max_tau0_over_l_max|}               & 15    & 2     & 2      &  2    & 2 \\\hline
\end{tabular}
\caption{ We show the precision parameters used~in~\texttt{CLASS} to compute the angular power spectrum of number counts fluctuations. For the observed spectrum $C_\ell^{\mathrm{obs,ij}}$  we use parameters yielding a high precision computation.  For the theoretical spectrum $C_\ell^{\mathrm{th,ij}}$, we adapt the precision parameters for each configuration in order to keep the error due to using lower precision parameters bound to $\Delta\chi^2\leq0.2$.  }
\label{tab:precision}
\end{table}

\begin{equation}
    \Delta \chi^2=\sum_{\ell=2}^{\ell_{\mathrm{max}}} (2\ell+1)f_{\mathrm{sky}} \left( \ln \frac{d_\ell^{\mathrm{th}}}{d_\ell^{\mathrm{obs}}}+ \frac{d_\ell^{\mathrm{mix}}}{d_\ell^{\mathrm{th}}} - N_{\mathrm{bins}}   \right), \label{eq:chi2}
\end{equation}
where~$d_\ell^\mathrm{A}=\det(C_\ell^{\mathrm{A,ij}})$ and~$d_\ell^{\mathrm{mix}}$ are calculated as~$d_\ell^{\mathrm{th}}$ but substituting in each term of the determinant one factor by~$C_\ell^{\mathrm{obs,ij}}$. We check that the possible error when calculating the~$C_\ell^{\mathrm{th,ij}}$ spectrum due to the lower precision parameters used is bound to be~$\Delta\chi^2\leq0.2$, which is achieved for the precision parameters of Table~\ref{tab:precision}. In Eq.~\eqref{eq:chi2} we choose~$\ell_{\mathrm{max}}=400$ in order to avoid strong contamination from non-linear scales. Thirdly, we speed up our analysis and break degeneracies by taking into consideration information about cosmological constraints from Subsection~\ref{subsection:cosmological-constraints}.  We use results for the case~\texttt{TTTEEE+lensing+BAO} and compute a covariance matrix from the chains. Then, covariance matrix~$\mathbf{C}$ for parameters~$\vec{\mathbf{x}}= (  \Omega_b h^2, \Omega_c h^2, n_s, \ln10^{10}A_s, \tau, H_0, w, \alpha)$ along with corresponding fiducial values in Table~\ref{Table:fiducial-model} are introduced in the forecast as a Gaussian prior. As for~$b_0$ and~$\sum m_\nu$ we use flat priors  given in Table~\ref{Table:flat-prior-bounds}. Thus, the~$\chi^2$ relative to the fiducial model including information from the prior reads
\begin{equation}
\Delta \chi^2=\sum_{\ell=2}^{\ell_{\mathrm{max}}} (2l+1)f_{\mathrm{sky}} \left( \ln \frac{d_\ell^{\mathrm{th}}}{d_\ell^{\mathrm{obs}}}+ \frac{d_\ell^{\mathrm{mix}}}{d_\ell^{\mathrm{th}}} -N_{\mathrm{bins}} \right) +\sum_{i,j}(x_i-x_i^{\mathrm{fid}})C_{ij}^{-1}(x_j-x_j^{\mathrm{fid}}), \label{eq:full_lkl}
\end{equation}
where~$\vec{\mathbf{x}}^{\mathrm{fid}}$ denotes parameters of the fiducial model in Table~\ref{Table:fiducial-model}.

\section{Results}
\label{section:results}

\subsection{Cosmological constraints}
\label{subsection:cosmological-constraints}

\begin{figure*}[http]
\centering
\includegraphics[scale=1.05]{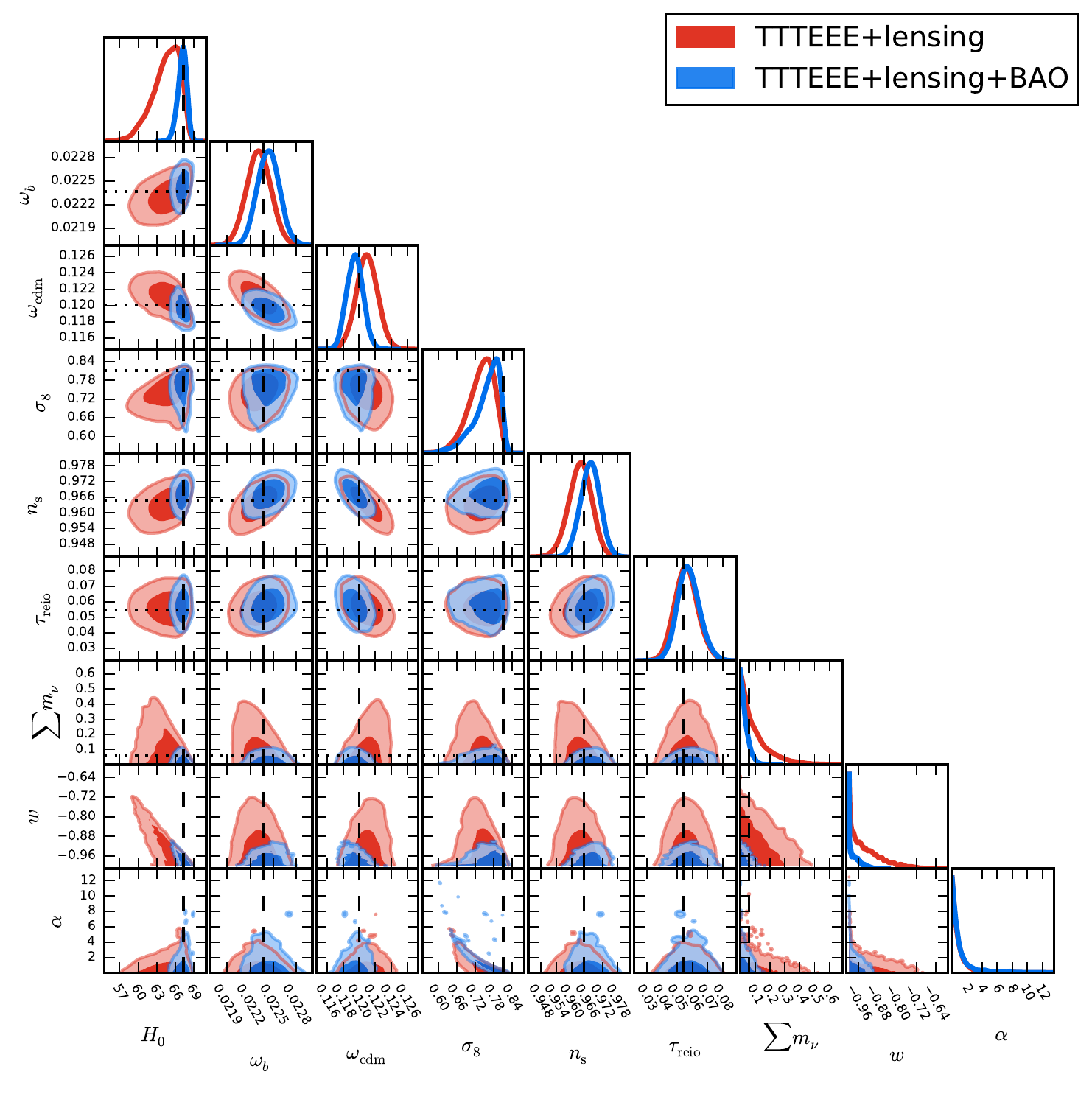}
\caption{The 1-D and 2-D posteriors for the cosmological parameters in the interacting model under consideration. Inner and outer contours correspond to $68\%$ and $95\%$ confidence regions, respectively.  Notation for data sets is explained in the main text, see Subsection~\ref{subsection:mcmc}.}
\label{Fig:constraints}
\end{figure*}

Here we constrain cosmological parameters of the interacting model discussed in Subsection~\ref{subsection:model}. Confidence contours for the posteriors of our analyses are shown in Figure~\ref{Fig:constraints} and the statistical information is summarised in Table~\ref{Table:statistical_info_parameters}. Vertical, dashed lines and horizontal, dotted lines in the triangle plot indicate the results reported by the Planck Collaboration (column \texttt{TT+TE+EE+lowE+lensing} in table 2 of Ref.~\cite{Aghanim:2018eyx}) for parameters in the $\Lambda$CDM   model. It is clear that constraints for parameters that are common to both interacting model and concordance model agree at the $2\sigma$ level.  

In this work we considered the sum of neutrino masses as a varying parameter. Table~\ref{Table:statistical_info_parameters} and Figure~\ref{Fig:constraints} allow us to conclude that the data cannot fully constrain the neutrino mass: we can only set an upper limit a bit weaker than recent results from the Karlsruhe Tritium Neutrino experiment~\cite{KATRIN:2021uub}. It is expected that upcoming surveys including information from non-linear scales might finally determine this parameter. In our analysis we also took into account a fluid with constant equation of state $w$ satisfying the condition $w > -1$ in order to avoid instabilities. Our results evidently prefer the value associated with a cosmological constant. With regard to the parameter controlling the strength of the DE-DM coupling, our results are in full agreement with a vanishing interaction in the dark sector. 

\begin{table}[http]
\centering
\begin{tabular}{@{}cc}
\hline
Parameter & Range \\
\hline
$b_0$ & $[0,3]$\\
$ \sum m_\nu $\,(eV) & $[0,5]$\\
\hline
\end{tabular}
\caption{Flat prior bounds used in the MCMC forecast.}
\label{Table:flat-prior-bounds}
\end{table}

\begin{table*}[t!]
\centering
\begin{tabular}{c c c}
\hline
Parameter  & \texttt{TTTEEE+lensing} & $\left\lbrace\dots\right\rbrace$\texttt{+BAO} \\
\hline
$\Omega_b h^2$ & $0.02231\pm 0.00016$ & $0.02244\pm 0.00014$ \\
$\Omega_c h^2$  & $0.1210\pm 0.0013$ & $0.1194^{-0.0011}_{+0.0010}$ \\
$H_0 \left(\frac{\text{km}}{\text{s}\cdot\text{Mpc}}\right) $ &  $64.52^{-1.39}_{+2.80}$ & $67.14^{-0.61}_{+0.84} $ \\
$ \ln 10^{10}A_s$  & $3.052^{-0.016}_{+0.014} $ & $3.052^{-0.015}_{+0.013}$ \\
$n_s$ &  $0.9634^{-0.0043}_{+0.0046}$ & $0.9673^{-0.0037}_{+0.0038}$\\
$\tau $  & $0.0561^{-0.0082}_{+0.0071}$ & $0.0576^{-0.0078}_{+0.0067}$\\
$\sum m_\nu $\,(eV)  & $<0.121$ & $<0.039$\\
$w $ &  $<-0.91$ & $ <-0.97 $ \\
$\alpha$  & $ < 1 $ & $ < 1 $ \\
$\sigma_8$ & $0.739^{-0.030}_{+0.051}$ & $0.757^{-0.020}_{+0.053}$ \\
\hline
\end{tabular}
\caption{Mean values and $68\%$ confidence limits on cosmological parameters for the elastic model. Here $\left\lbrace\dots\right\rbrace$ stands for the inclusion of data from column on the left.}
\label{Table:statistical_info_parameters}
\end{table*}

\subsection{Forecast}
\label{subsection:forecast}

In this section we present forecasts for a EUCLID-like galaxy survey. In order to estimate the relevance of lensing convergence when modelling number count fluctuations we carry out two kinds of analyses. Firstly, in our analysis ``with lensing'' we model angular power spectrum of number count fluctuations by taking into consideration matter perturbations, redshift space distortions, and lensing convergence. Secondly, in the analysis ``without lensing'' we only consider matter perturbations and redshift space distortions. Concerning the galaxy distribution we also take into account different configurations. Since the results might depend on the number of redshift bins~$N_{\mathrm{bins}}$ used, we perform the analyses with two configurations using~$N_{\mathrm{bins}} = 5,10$. Regarding the shape of the galaxy distribution we study two possibilities, namely, Gaussian and top-hat. In both cases, we make sure the number of galaxies per redshift bin is evenly distributed.

\subsubsection{Top-hat: 5 redshift bins}
\label{sec:5bins_top_hat}

\begin{figure*}[t!]
\centering
\includegraphics[scale=1.]{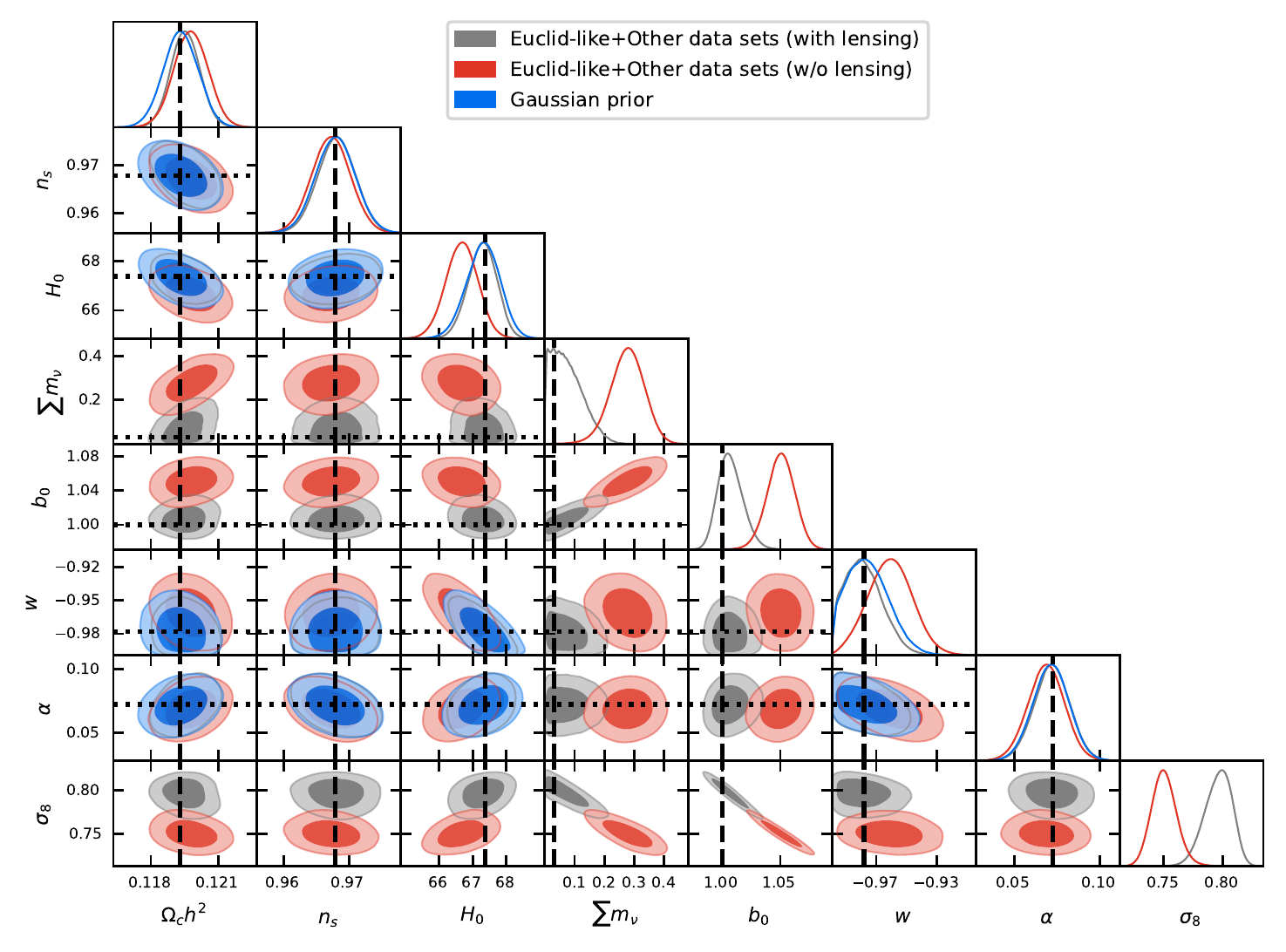}
\caption{The 1-D and 2-D posteriors for the cosmological survey and model  parameters. Here the analysis uses a 5 redshift bins top-hat galaxy density distribution. Gray (red) contours indicate results when lensing convergence is included (neglected), whereas in blue we show the Gaussian prior distribution. Black, dashed, vertical and black, dotted, horizontal lines indicate the values of the fiducial model.}
\label{Fig:forecast-tophat-5-bins}
\end{figure*}

As we can see in Table~\ref{Table:forecast-tophat-5-bins} and in Figure~\ref{Fig:forecast-tophat-5-bins}, in the analysis consistently including lensing convergence  all the parameters are inside the $1\sigma$ region giving no shift with respect to the fiducial cosmology. When we neglect lensing convergence, the background parameters $w$ and $H_0$ have a $\sim1-2\sigma$ shift with respect to the fiducial cosmology both in the mean and best fit values. It is important to remark that  the shift on those parameters occurs despite using a Gaussian prior on them. The bias parameter $b_0$ and the neutrino mass $\sum m_\nu$, which are not affected by the Gaussian prior, have a $\sim 4\sigma$ shift with respect to the fiducial cosmology. It is worth saying that when we neglect lensing convergence we obtain a spurious  detection of the neutrino mass, whereas  when consistently including lensing convergence  we are only able to put an upper constraint. These results are in agreement with Ref.~\cite{Cardona:2016qxn} and should be considered as a warning of how some approximations can lead to biased results in analyses of upcoming galaxy surveys.

We must highlight the fact that all these shifts are uncorrelated to the model parameter $\alpha$ as it has no shift when we do not consider the lensing contributions. Nonetheless, we can see in Figure~\ref{Fig:forecast-tophat-5-bins} that the derived parameter $\sigma_8$ also shows a shift to lower values, that is, less structure in the Universe when we neglect lensing  convergence in the analysis. The amplitude of matter fluctuations is key in all the momentum transfer coupling (see Refs.~\cite{Figueruelo:2021elm,Linton:2021cgd,BeltranJimenez:2020qdu,Simpson:2010vh,Jimenez:2021ybe,Pourtsidou:2016ico, Jimenez:2020ysu,Chamings:2019kcl}) as a non-vanishing interaction  tends to lower  $\sigma_8$. Since our analyses do not show any shift in the coupling parameter, we understand the lower value of $\sigma_8$ as a consequence of neglecting lensing convergence when modelling number counts: there is a degeneracy with both $b_0$ and $\sum m_\nu$, parameters heavily shifted with respect to the fiducial values. We must bear in mind that $\sigma_8$ has no fiducial value: $\sigma_8$ is derived from our samples. Consequently, shifts on $\sigma_8$ might be due to a mixture of effects (e.g., shifts on other cosmological parameters determining $\sigma_8$).

\begin{table}[t!]
  \centering
  \begin{tabular}{@{}cccccc}
    \hline
    \multicolumn{6}{c}{i) Consistently including lensing: $\Delta \chi^2 = 0$} \\
    \hline
    Parameter & Mean & Best fit & $\sigma$ &\hspace{-0.52cm} shift: Mean & Best fit \\
    \hline
    $\Omega_b h^2$ & $0.02244 $ & $0.02248 $ & $ 0.00011$ &  \quad$0.2\sigma$ & $ 0.1\sigma$ \\
    $\Omega_{c} h^2$ & $0.1195 $ & $ 0.1193$ & $ 0.0006$  &  \quad$0.3\sigma$ & $0.1\sigma$ \\
    $n_s$      & $0.9681 $ & $ 0.9691$ & $ 0.0029$ &  \quad$0.1\sigma$ & $0.4\sigma$ \\
    $\ln10^{10}A_s$       & $3.045$&  $3.049$ &$ 0.013$ &  \quad$<0.1\sigma$ & $ 0.4\sigma$ \\
    $\tau$ & $ 0.0544$ &  $ 0.0570$ & $ 0.0059$ & \quad$<0.1\sigma$ & $0.5\sigma$ \\
    $H_0\left(\frac{\text{km}}{\text{s}\cdot\text{Mpc}}\right)$ & $ 67.29$ & $67.40 $ & $ 0.42$ &  \quad$0.2 \sigma$ & $ <0.1 \sigma$ \\
    $w$ & $ -0.9781$ & $ -0.9794$ & $ 0.01$ & \quad$<0.1 \sigma$ & $0.1 \sigma$ \\
    $b_0$ & $1.007 $ & $0.999 $ & $ 0.011$ & \quad$0.7\sigma$ & $0.1 \sigma$ \\
    $\sum m_{\nu}$\,(eV)  & $ 0.0758$ & $ 0.0357$ & $ 0.05$ & \quad $ 0.9\sigma$ & $<0.1 \sigma$ \\
    $\alpha$ & $ 0.0718$ & $0.0708 $ & $ 0.01$ & \quad$<0.1\sigma$ & $0.1\sigma$ \\	
    \end{tabular}
  \begin{tabular}{@{}cccccc}
    \hline
    \multicolumn{6}{c}{ii) Neglecting lensing: $\Delta \chi^2 = 1636$} \\
    \hline
    Parameter & Mean & Best fit & $\sigma$ &\hspace{-0.52cm} shift: Mean & Best fit \\
    \hline
    $\Omega_b h^2$ & $0.02240 $ & $ 0.02238$ & $ 0.00012$ &  \quad$0.5\sigma$ & $ 0.7\sigma$ \\
    $\Omega_{c} h^2$ & $0.1198 $ & $ 0.1196$ & $ 0.0008$  &  \quad$0.6\sigma$ & $0.3\sigma$ \\
    $n_s$      & $ 0.9673$ & $ 0.9684$ & $ 0.0029$ &  \quad$0.2\sigma$ & $0.2\sigma$ \\
    $\ln10^{10}A_s$       & $3.0398$&  $3.041$ &$ 0.014$ &  \quad$0.3\sigma$ & $ 0.2\sigma$ \\
    $\tau$ & $ 0.0523$ &  $ 0.0516$ & $ 0.0060$ & \quad$0.3\sigma$ & $0.4\sigma$ \\
    $H_0\left(\frac{\text{km}}{\text{s}\cdot\text{Mpc}}\right)$ & $ 66.67$ & $66.72 $ & $ 0.47$ &  \quad$ 1.5\sigma$ & $ 1.4 \sigma$ \\
    $w$ & $ -0.9608$ & $ -0.9613$ & $ 0.014$ & \quad$ 1.2\sigma$ & $ 1.2\sigma$ \\
    $b_0$ & $ 1.0503$ & $ 1.0499$ & $ 0.012$ & \quad$4.2\sigma$ & $4.2 \sigma$ \\
    $\sum m_{\nu}$\,(eV)  & $ 0.276$ & $ 0.273$ & $ 0.056$ & \quad $ 4.3\sigma$ & $4.3 \sigma$ \\
    $\alpha$ & $ 0.0686$ & $0.0683 $ & $ 0.011$ & \quad$0.3\sigma$ & $0.4\sigma$ \\

\hline
\end{tabular}
\caption{The statistical results and the respective shifts with respect to the fiducial cosmology when we consider all the contributions to the angular power spectrum of number counts fluctuation (up) and when  we neglect lensing convergence (down) for a 5 redshift bins top-hat galaxy density distribution.}
\label{Table:forecast-tophat-5-bins}
\end{table}

\subsubsection{Gaussian: 5 redshift bins}

\begin{figure*}[t!]
\centering
\includegraphics[scale=1.]{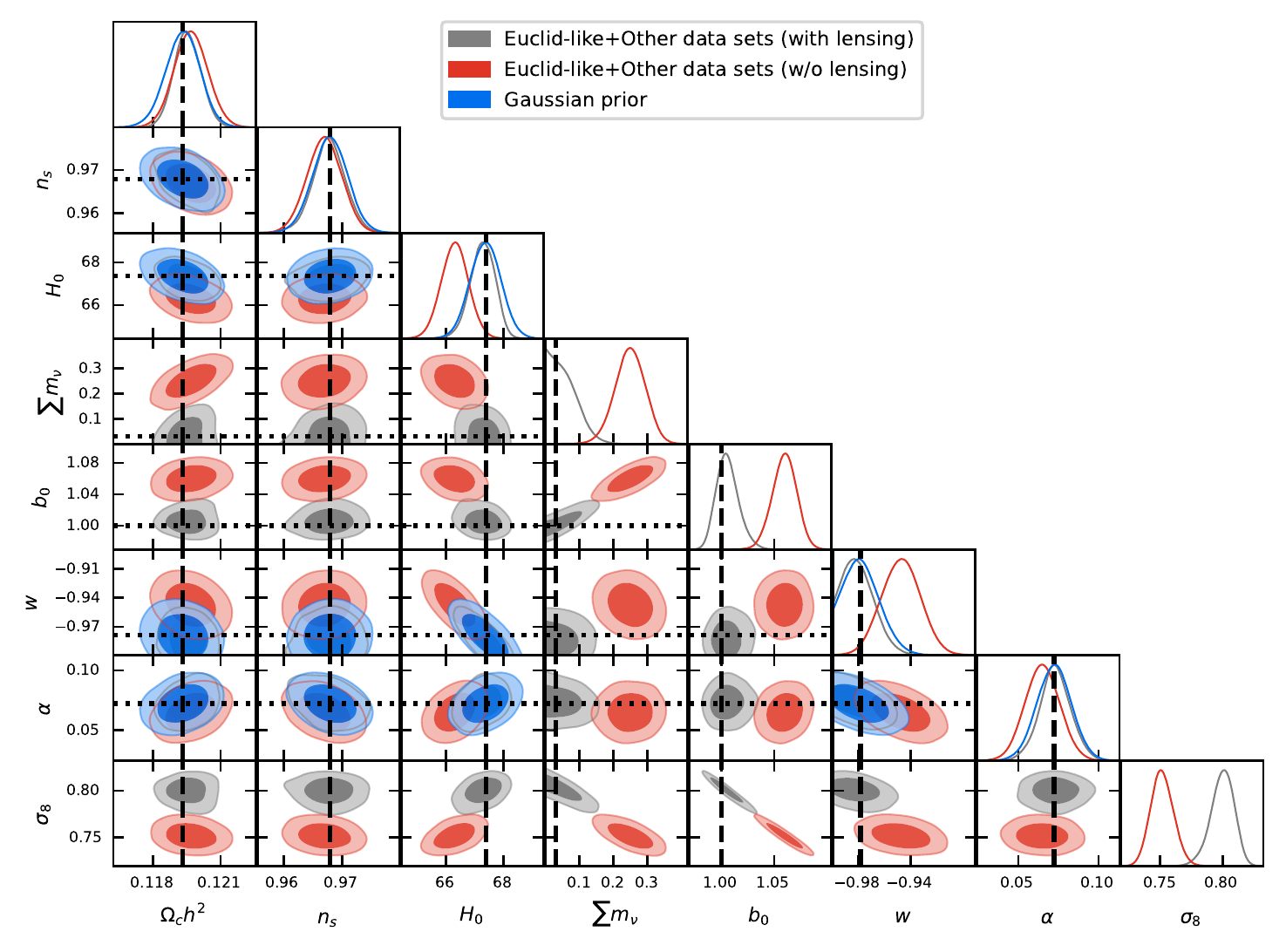}
\caption{The 1-D and 2-D posteriors or the cosmological, survey and model parameters. Here the analysis uses a 5 redshift bins Gaussian galaxy density distribution. Gray (red) contours indicate results when lensing convergence is included (neglected), whereas in blue we show the Gaussian prior distribution. Black, dashed, vertical and black, dotted, horizontal lines indicate the values of the fiducial model.} 
\label{Fig:forecast-Gaussian-5-bins}
\end{figure*}

We carry out a forecast taking into consideration a Gaussian galaxy distribution with $5$ redshift bins. Results are depicted in Figure~\ref{Fig:forecast-Gaussian-5-bins} and the statistical information shown in Table~\ref{Table:forecast-Gaussian-5-bins}. Whereas gray contours indicate an analysis that consistently includes lensing convergence when modelling number counts fluctuations, red contours show the results when we disregard lensing convergence. The consistent analysis shows that we are able to recapture the fiducial values of the cosmological parameters represented by black, dashed and black, dotted lines in Figure~\ref{Fig:forecast-Gaussian-5-bins}. The situation is rather different when we do not include lensing convergence. In this case we observe mild ($\approx 2\sigma$) biased constraints on the Hubble constant $H_0$ and the equation of state $w$, and severe ($\approx 5\sigma$) shifts on the neutrino mass $\sum m_\nu$ and the bias amplitude $b_0$. The latter is clearly explained by the $\sum m_\nu-b_0$ degeneracy (see Figure~\ref{Fig:forecast-Gaussian-5-bins}). Here we also see the shift to lower values of $\sigma_8$ when we neglect the lensing convergence contribution, not depending on the value of $\alpha$. These results are in good agreement with Subsection~\ref{sec:5bins_top_hat} where a top-hat galaxy distribution was used. 

\begin{table}[t!]
  \centering
  \begin{tabular}{@{}cccccc}
    \hline
    \multicolumn{6}{c}{i) Consistently including lensing: $\Delta \chi^2 = 0$} \\
    \hline
    Parameter & Mean & Best fit & $\sigma$ &\hspace{-0.52cm} shift: Mean & Best fit \\
    \hline
    $\Omega_b h^2$ & $0.02244$ & $0.02247 $ & $0.00011 $ &  \quad$0.3\sigma$ & $ <0.1\sigma$ \\
    $\Omega_c h^2$ & $0.1195 $ & $0.1193 $ & $0.0006 $ &  \quad$0.4\sigma$ & $<0.1\sigma$ \\
    $n_s$      & $0.9678 $ & $0.9679 $ & $0.0028 $ &  \quad$0.1\sigma$ & $ <0.1\sigma$ \\
    $\ln10^{10}A_s$ & $3.043 $ & $3.044$ & $0.013 $ &  \quad$0.1\sigma$ & $ <0.1\sigma$ \\
    $H_0\left(\frac{\text{km}}{\text{s}\cdot\text{Mpc}}\right)$ & $67.27$ & $66.38$ & $0.4$ &  \quad$0.3\sigma$ & $ <0.1\sigma$ \\
    $\sum m_{\nu}$\,(eV)  & $0.06$ & $0.03$ & $0.04$ & \quad $ 0.6\sigma$ & $ <0.1\sigma$ \\
    $b_0$ & $1.006$ & $1.000$ & $0.011$ & \quad$0.5\sigma$ & $<0.1\sigma$ \\
    $w$ & $-0.98$ & $-0.98$ & $0.01$ & \quad$<0.1\sigma$ & $<0.1\sigma$ \\
    $\tau$ & $0.0539$ & $0.0543$ & $0.0056$ & \quad$0.1\sigma$ & $<0.1\sigma$ \\
    $\alpha$ & $0.0730$ & $0.0723$ & $0.0096$ & \quad$0.1\sigma$ & $<0.1\sigma$ \\	
  \end{tabular}
  \begin{tabular}{@{}cccccc}
    \hline
    \multicolumn{6}{c}{ii) Neglecting lensing: $\Delta \chi^2 = 1835$} \\
    \hline
    Parameter & Mean & Best fit & $\sigma$ & \hspace{-0.52cm} shift: Mean & Best fit \\
    \hline
    $\Omega_b h^2$ & $0.02240$ & $0.02240 $ & $0.00012 $ &  \quad$0.6\sigma$ & $0.6\sigma$ \\
    $\Omega_c h^2$ & $0.1197$ & $0.1196$ & $0.0007$ &  \quad$0.5\sigma$ & $0.4\sigma$ \\
    $n_s$      & $0.9670$ & $0.9680$ & $0.0030$ &  \quad$0.3\sigma$ & $<0.1\sigma$ \\
    $\ln10^{10}A_s$ & $ 3.039 $ & $3.035 $ & $ 0.014 $ &  \quad$0.4\sigma$ & $0.6\sigma$ \\
    $H_0\left(\frac{\text{km}}{\text{s}\cdot\text{Mpc}}\right)$      & $66.31$ & $66.42$ & $0.47$ &  \quad$2.3\sigma$ & $2.0\sigma$ \\
    $\sum m_{\nu}$\,(eV)  & $0.25$ & $0.26$ & $0.04$ &  \quad$4.9\sigma$ & $5.1\sigma$ \\
    $b_0$ & $1.060$ & $1.064$ & $0.012$ & \quad$5.2\sigma$ & $5.5\sigma$\\
    $w$ & $-0.95$ & $-0.95$ & $0.01$ & \quad$2.2\sigma$ & $2.0\sigma$ \\
    $\tau$ & $0.0516$ & $0.0493$ & $0.061$ & \quad$0.4\sigma$ & $0.8\sigma$ \\
    $\alpha$ & $0.0650$ & $0.0689$ & $0.0107$ & \quad$0.7\sigma$ & $0.3\sigma$ \\	
\hline
\end{tabular}
\caption{The statistical results and the respective shifts to the fiducial cosmology when we consider all the contributions to the angular power spectrum of number counts fluctuation (up) and when  we neglect lensing convergence (down) for a 5 bins Gaussian galaxy density distribution.}
\label{Table:forecast-Gaussian-5-bins}
\end{table}

\subsubsection{Top-hat: 10 redshift bins}
\label{sec:10bins_top_hat}

\begin{figure*}[t!]
\centering
\includegraphics[scale=1.06]{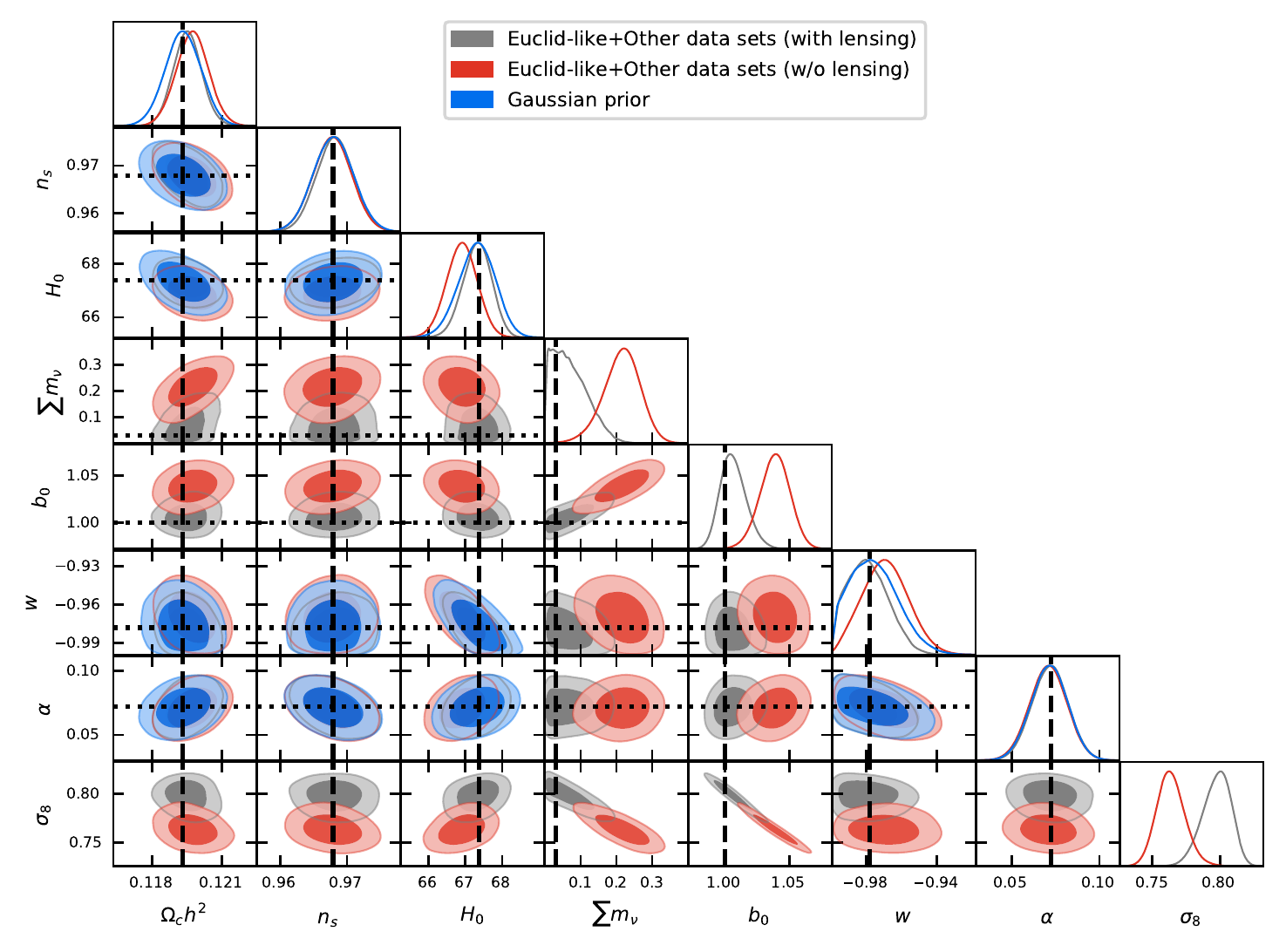}
\caption{The 1-D and 2-D posteriors for the cosmological, survey and model parameters. Here the analysis uses a 10 redshift bins top-hat galaxy density distribution. Gray (red) contours indicate results when lensing convergence is included (neglected), whereas in blue we show the Gaussian prior distribution. Black, dashed, vertical and black, dotted, horizontal lines indicate the values of the fiducial model. }
\label{Fig:forecast-tophat-10-bins} 
\end{figure*}

Here we study a top-hat distribution of galaxies as in Subsection~\ref{sec:5bins_top_hat}, but now using $10$ redshift bins instead. Results are shown in Table~\ref{Table:forecast-tophat-10-bins} and Figure~\ref{Fig:forecast-tophat-10-bins}. If we include all relevant contributions when modelling angular power spectrum of number count fluctuations (i.e., density, redshift space distortions, and lensing convergence), we see no significant discrepancies ($<1\sigma$) in the values of cosmological parameters with respect to the fiducial cosmology. As we see above in the case with $5$ redshift bins, neglecting lensing convergence in the analysis also has an important effect in the determination of cosmological parameters. Although here shifts are slightly smaller than those found in Subsection~\ref{sec:5bins_top_hat}, we still find a $3\sigma$ difference with respect to the fiducial values in the parameters $\sum m_\nu$ and $b_0$ as well as a shift in the amplitude of matter perturbations $\sigma_8$, as in the 5 redshift bins case. Concerning the parameter $\alpha$ governing the DM-DE interaction we find it is insensitive to neglecting lensing convergence. From Subsection~\ref{sec:5bins_top_hat} and the current analysis we conclude that our results do not show a strong dependence on the number of redshift bins, so that neglecting lensing convergence when modelling number counts fluctuations can lead to biased cosmological constraints regardless of the configuration for the galaxy distribution. Nevertheless, enlarging the number of bins seems to reduce the shifts when we neglect the lensing contributions to the computation of the angular power spectrum of number counts fluctuation. 

\begin{table}[t!]
  \centering
  \begin{tabular}{@{}cccccc}
    \hline
    \multicolumn{6}{c}{i) Consistently including lensing: $\Delta \chi^2 = 0$} \\
    \hline
    Parameter & Mean & Best fit & $\sigma$ &\hspace{-0.52cm} shift: Mean & Best fit \\
    \hline
    $\Omega_b h^2$ & $0.02245$ & $0.02248$ & $0.00011$ &  \quad$0.2\sigma$ & $ 0.2\sigma$ \\
    $\Omega_{c} h^2$  & $0.1195$ & $0.1193$ &  $0.0006 $ &  \quad$0.3\sigma$ & $0.1\sigma$ \\
    $n_s$             & $0.9682$ & $ 0.9690$ & $ 0.0028$ &  \quad$0.1\sigma$ & $0.4\sigma$ \\
    $\ln10^{10}A_s$       & $3.045$&  $3.048$ &$ 0.013$ &  \quad$<0.1\sigma$ & $ 0.4\sigma$ \\
    $\tau$     & $0.0544$ & $ 0.0556$  & $0.0058$ & \quad$<0.1\sigma$ & $0.3\sigma$ \\
    $H_0\left(\frac{\text{km}}{\text{s}\cdot\text{Mpc}}\right)$ & $67.32$ & $67.40$  & $0.39$ &  \quad$ 0.2\sigma$ & $<0.1  \sigma$ \\
    $w$ & $ -0.9788$ & $ -0.9792$ & $ 0.011$ & \quad$<0.1\sigma$ & $ <0.1\sigma$ \\
    $b_0$ & $1.006 $ & $1.004 $ & $0.010 $ & \quad$0.6\sigma$ & $ 0.4\sigma$ \\
    $\sum m_{\nu}$\,(eV)  & $ 0.0705$ & $0.0612 $ & $ 0.048$ & \quad $ 0.8\sigma$ & $ 0.6\sigma$ \\
    $\alpha$ & $0.0718 $ & $ 0.0685$ & $0.010 $ & \quad$<0.1\sigma$ & $0.4\sigma$ \\	
    \end{tabular}
  \begin{tabular}{@{}cccccc}
    \hline
    \multicolumn{6}{c}{ii) Neglecting lensing: $\Delta \chi^2 = 1988$} \\
    \hline
    Parameter & Mean & Best fit & $\sigma$ &\hspace{-0.52cm} shift: Mean & Best fit \\
    \hline
    $\Omega_b h^2$ & $0.02240$ & $0.02237$ & $0.00012$ &  \quad$0.5\sigma$ & $ 0.8\sigma$ \\
    $\Omega_{c} h^2$ & $0.1197$ & $0.1198$ & $0.0007$ &  \quad$0.5\sigma$ & $0.7\sigma$ \\
    $n_s$      & $0.9678$ & $0.9670 $ & $0.0029$  &  \quad$<0.1\sigma$ & $0.3\sigma$ \\
    $\ln10^{10}A_s$       & $3.0378$&  $3.036$ &$ 0.014$ &  \quad$0.3\sigma$ & $ 0.2\sigma$ \\
    $\tau$ & $0.0516$ & $ 0.0511$  & $ 0.0061$ & \quad$0.4\sigma$ & $0.5\sigma$ \\
    $H_0\left(\frac{\text{km}}{\text{s}\cdot\text{Mpc}}\right)$ & $ 66.90$ & $ 67.01$ & $ 0.42$ &  \quad$1.1 \sigma$ & $ 0.9 \sigma$ \\
    $w$ & $-0.9691$ & $ -0.9728$ & $0.013 $ & \quad$ 0.7\sigma$ & $ 0.4\sigma$ \\
    $b_0$ & $1.0384 $  & $1.0376 $ & $0.012 $ & \quad$3.3\sigma$ & $3.2\sigma$ \\
    $\sum m_{\nu}$\,(eV)  & $0.214$ & $ 0.206$ & $0.052 $ & \quad $ 3.5\sigma$ & $3.4\sigma$ \\
    $\alpha$ & $0.0708$ & $ 0.0730$ & $0.010$ & \quad$0.1\sigma$ & $<0.1\sigma$ \\	
    
\hline
\end{tabular}
\caption{The statistical results and the respective shifts with respect to the fiducial cosmology when we consider all the contributions to the angular power spectrum of number counts fluctuation (up) and when  we neglect lensing convergence (down) for a 10 redshift bins top-hat galaxy density distribution. }
\label{Table:forecast-tophat-10-bins}
\end{table}

\subsubsection{Gaussian: 10 redshift bins}
 
\begin{figure*}[t!]
\centering
\includegraphics[scale=1.]{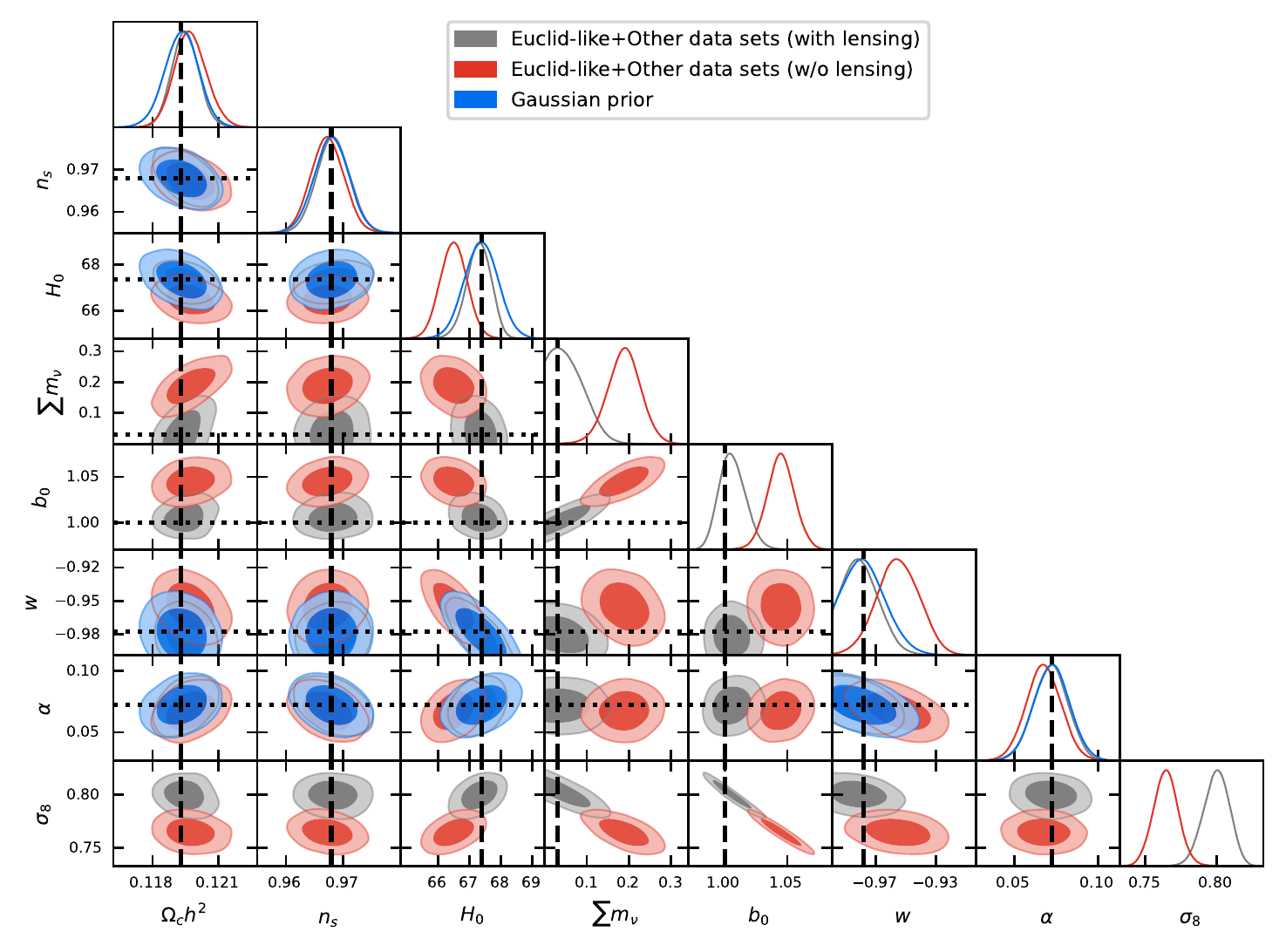}
\caption{The 1-D and 2-D posteriors for the cosmological survey and model parameters. Here the analysis uses a 10 redshift bins Gaussian galaxy density distribution. Gray (red) contours indicate results when lensing convergence is included (neglected), whereas in blue we show the Gaussian prior distribution. Black, dashed, vertical and black, dotted, horizontal lines indicate the values of the fiducial model.
}
\label{Fig:forecast-Gaussian-10-bins}
\end{figure*}

Results for our forecast using $10$ Gaussian redshift bins are shown in Figure~\ref{Fig:forecast-Gaussian-10-bins} and Table~\ref{Table:forecast-Gaussian-10-bins}. Parameters $\Omega_b h^2$, $\Omega_c h^2$, $n_s$, $\ln 10^{10}A_s$, $\tau$, $\alpha$ do not change significantly with respect to their prior distribution. The analysis neglecting lensing convergence in the modelling of number counts fluctuations shows mild shifts ($1-2\sigma$) with respect to the fiducial model in the Hubble constant $H_0$ and the DE equation of state $w$. Our analysis also shows important ($\approx 4\sigma$) biased constraints on the neutrino mass and the bias amplitude. Interestingly, when consistently including lensing convergence in the analysis, constraints are in very good agreement with the fiducial values. As in the previous cases, we find a shift of $\sigma_8$ when we neglect the lensing convergence contribution. 
 
\begin{table}[t!]
  \centering
  \begin{tabular}{@{}cccccc}
    \hline
    \multicolumn{6}{c}{i) Consistently including lensing: $\Delta \chi^2 = 0$} \\
    \hline
    Parameter & Mean & Best fit & $\sigma$ &\hspace{-0.52cm} shift: Mean & Best fit \\
    \hline
    $\Omega_b h^2$ & $0.02245$ & $0.02242 $ & $0.00011 $ &  \quad$0.2\sigma$ & $0.5\sigma$ \\
    $\Omega_c h^2$ & $0.1195 $ & $0.1195 $ & \quad$0.0006 $ &  \quad$0.3\sigma$ & $0.4\sigma$ \\
    $n_s$      & $0.9682 $ & $0.9687 $ & $0.0029 $ &  \quad$0.1\sigma$ & $ 0.3\sigma$ \\
    $\ln10^{10}A_s$ & $3.044 $ & $3.046$ & $0.013 $ &  \quad$<0.1\sigma$ & $ 0.1\sigma$ \\
    $H_0\left(\frac{\text{km}}{\text{s}\cdot\text{Mpc}}\right)$ & $67.29$ & $67.26$ & $0.4$ &  \quad$0.2\sigma$ & $ 0.3\sigma$ \\
    $\sum m_{\nu}$\,(eV)  & $0.06$ & $0.06$ & $0.04$ & \quad $ 0.7\sigma$ & $ 0.8\sigma$ \\
    $b_0$ & $1.006$ & $1.005$ & $0.010$ & \quad$0.6\sigma$ & $0.5\sigma$ \\
    $w$ & $-0.980$ & $-0.983$ & $0.011$ & \quad$<0.1\sigma$ & $0.2\sigma$ \\
    $\tau$ & $0.0540$ & $0.0549$ & $0.0057$ & \quad$<0.1\sigma$ & $0.1\sigma$ \\
    $\alpha$ & $0.0717$ & $0.0715$ & $0.0102$ & \quad$0.1\sigma$ & $0.1\sigma$ \\	
  \end{tabular}
  \begin{tabular}{@{}cccccc}
    \hline
    \multicolumn{6}{c}{ii) Neglecting lensing: $\Delta \chi^2 = 2435$} \\
    \hline
    Parameter & Mean & Best fit & $\sigma$ & \hspace{-0.52cm} shift: Mean & Best fit \\
    \hline
    $\Omega_b h^2$ & $0.02239$ & $0.02245 $ & $0.00012 $ &  \quad$0.6\sigma$ & $0.1\sigma$ \\
    $\Omega_c h^2$ & $0.1197$ & $0.1196$ & $0.0007$ &  \quad$0.6\sigma$ & $0.4\sigma$ \\
    $n_s$      & $0.9672$ & $0.9681$ & $0.0029$ &  \quad$0.2\sigma$ & $0.1\sigma$ \\
    $\ln10^{10}A_s$ & $ 3.037 $ & $3.035 $ & $ 0.014 $ &  \quad$0.5\sigma$ & $0.7\sigma$ \\
    $H_0\left(\frac{\text{km}}{\text{s}\cdot\text{Mpc}}\right)$      & $66.49$ & $66.66$ & $0.42$ &  \quad$2.1\sigma$ & $1.7\sigma$ \\
    $\sum m_{\nu}$\,(eV)  & $0.19$ & $0.20$ & $0.04$ &  \quad$4.0\sigma$ & $4.1\sigma$ \\
    $b_0$ & $1.045$ & $1.048$ & $0.011$ & \quad$4.1\sigma$ & $4.4\sigma$\\
    $w$ & $-0.955$ & $-0.961$ & $0.014$ & \quad$1.8\sigma$ & $1.4\sigma$ \\
    $\tau$ & $0.0511$ & $0.0513$ & $0.0060$ & \quad$0.5\sigma$ & $0.5\sigma$ \\
    $\alpha$ & $0.0675$ & $0.0695$ & $0.0104$ & \quad$0.5\sigma$ & $0.3\sigma$ \\	
\hline
\end{tabular}
\caption{The statistical results and the respective shifts to the fiducial cosmology when we consider all the contributions to the angular power spectrum of number counts fluctuation (up) and when  we neglect lensing convergence (down) for a 10 bins Gaussian galaxy density distribution.}
\label{Table:forecast-Gaussian-10-bins}
\end{table}

\section{Discussion}
\label{section:discussion}

In this paper we considered a cosmological model where dark energy and dark matter are allowed to interact with each other via a momentum transfer. While this elastic coupling mainly affects the evolution of perturbations on small scales, the background evolution remains unchanged with respect to a model with a vanishing elastic interaction $\alpha$. Non-linear scales will play a part in analyses of upcoming galaxy surveys, hence it is crucial to investigate possible systematic effects that could hinder the accurate, precise determination of cosmological parameters.

Here, we utilised recent data sets and computed cosmological constraints (see Figure~\ref{Fig:constraints} and Table~\ref{Table:statistical_info_parameters}). Our results for the cosmological parameters in common with the standard cosmological model $\Lambda$CDM are in good agreement with the baseline analysis by the Planck Collaboration. Concerning additional parameters, note that constraints on the neutrino mass $\sum m_\nu$ hit the lower limit of the prior and we can only set an upper bound which agrees with recent direct neutrino mass measurements of the Karlsruhe Tritium Neutrino experiment~\cite{KATRIN:2021uub}. The DE equation of state $w$ presents a similar behaviour to the neutrino mass: there is good agreement with the value for a cosmological constant $w=-1$, that we set as our lower limit in the parameters space. The parameter $\alpha$ governing the elastic interaction is, in all cases, compatible with an uncoupled dark sector. Previous works, using the same data sets as we do here, have also found an upper bound for the coupling $\alpha$ in good agreement with our results (see Ref.~\cite{Figueruelo:2021elm} and Ref.~\cite{Jimenez:2021ybe} where extra radiation is added via a free $N_{eff}$ parameter). However, when the Sunyaev–Zeldovich likelihood is included in the MCMC analysis, a clear detection of the interaction can be found~\cite{Asghari:2019qld,Figueruelo:2021elm,Jimenez:2021ybe}. We also confirm that introducing an elastic scattering in the dark sector does not alleviate the discrepancy in the Hubble constant, as the interaction does not modify the background evolution. The interaction however has a direct impact in the clustering rate of dark matter: for \texttt{TTTEEE+lensing} we obtain a $\sigma_8$ value $\approx 1.4\sigma$ lower than in the standard model and in excellent agreement ($\approx 0.1\sigma$) with determinations from low red-shift probes.\footnote{Uncertainties added in quadrature. Planck Collaboration baseline result is $\sigma_8=0.8111\pm 0.0060$ \cite{Aghanim:2018eyx}. Recent DES value $\sigma_8=0.733^{+0.039}_{-0.049}$ \cite{DES:2021wwk}.}    

Cosmological constraints on key parameters such as the neutrino mass are expected to be greatly improved by the advent of galaxy surveys like EUCLID or LSST. While current analyses disregard details in the modelling of number counts~\cite{PhysRevLett.122.171301}, upcoming surveys demand a more careful treatment if biased constraints are to be avoided. A number of works have shown that relativistic effects cannot be neglected any longer and the relevance of lensing convergence in analyses of forthcoming galaxy surveys has been assessed in different cosmological models~\cite{Euclid:2021rez,PhysRevD.97.023537,Cardona:2016qxn,PhysRevD.83.123514,Duncan:2013haa}. As a result, we now know that neglecting lensing convergence in analyses could lead to biased constraints in the DE equation of state $w$, non-Gaussianity $f_{\rm{NL}}$, neutrino mass $\sum m_\nu$, and Modified-Gravity parameters. Here we considered a cosmological model where $w$, $\sum m_\nu$, as well as $\alpha$ (a parameter governing a possible elastic interaction in the dark sector) are varying parameters. Since the elastic model mainly affects the evolution of perturbations on small scales where upcoming surveys will add valuable information, we carried out forecasts for an EUCLID-like galaxy survey to assess how relevant lensing convergence will be in this context.

In order to make a realistic analysis we took into account information from the cosmological constraints computed in Subsection~\ref{subsection:cosmological-constraints} and performed MCMC forecasts using different configurations for the galaxy distribution. In general we expect our results to depend on the number of bins in which we split the galaxy survey as well as in the shape of the galaxy redshift bins. On the one hand, the impact of lensing convergence on the determination of cosmological parameters will have a bigger impact for a small number of wider redshift bins. In wider redshift bins radial correlations are suppressed and the constraining power comes mainly from transverse correlations where lensing convergence plays an important role. On the other hand, if we split the survey in a bigger number of thinner redshift bins, we expect an increase only in the number of modes dominated by density and RSD (but not the ones induced by lensing convergence); in such a galaxy survey configuration the impact of lensing convergence would not be as important as in the previous case. The optimal binning for galaxy clustering has been studied in detail in~\cite{Euclid:2021osj} and it turns out to be similar to the cases used here with $10$ bins. We do not expect our results to change for a greater number of redshift bins, since the analysis is limited by photo-z precision in upcoming galaxy surveys.

Indeed, Tables~\ref{Table:forecast-tophat-5-bins}-\ref{Table:forecast-Gaussian-10-bins} show slightly different results for biased parameters depending on the shape of the redshift bins (i.e., Gaussian and top-hat) as well as the number of bins in which we divide the galaxy survey. We can see that considering cases with equal number of bins (i.e., either $5$ or $10$), Gaussian redshift bins tend to give greater shifts in the cosmological parameters than top-hat redshift bins. Concerning the dependence of our results with the number of redshift bins, we can see that whatever the shape of the redshift bins, a galaxy survey divided into smaller, wider bins  enhances the effect of lensing convergence and therefore leads to slightly bigger shifts in cosmological parameters as we explained above.

\section{Conclusions}
\label{section:conclusions}

It might be possible that Dark Energy and Dark Matter interact with each other not only through gravity. Here we investigated a cosmological model where the dark sector is coupled via a momentum transfer that changes the evolution of perturbations mainly on small scales while having no impact on the background evolution.

We computed cosmological constraints by using recent measurements of CMB and BAO.  Our results show good agreement with a vanishing interaction in the dark sector. Previous works found that also taking into account SZ data yields a $\gtrsim3\sigma$ detection; these works also found a low $\chi^2$ which according to BIC and AIC criteria would favour the interaction model. Our analysis shows that due to an enlarged parameter space these kinds of models featuring DM-DE interaction might be disfavoured when carrying out Bayesian model comparison with respect to the standard model. Nevertheless, the presence of a momentum transfer in the dark sector alleviates the discrepancy in the amplitude of matter fluctuations found in analyses with the standard model $\Lambda$CDM.

Forthcoming galaxy surveys will exploit clustering information on small scales where elastic interactions in the dark sector and relativistic effects such as lensing convergence might play a part. In this work we also performed forecasts for an EUCLID-like galaxy survey in order to assess the relevance lensing convergence might have when modelling number counts in the context of a dark sector momentum transfer. We confirmed that lensing convergence must be taken into consideration in analyses of upcoming galaxy surveys. If we decide to ignore lensing convergence, we might face precise but innacurate (biases ranging from $1\sigma$ up to $5\sigma$) determination of important cosmological parameters such as the neutrino mass $\sum m_\nu$, the Hubble constant $H_0$, galaxy bias $b_0$, dark energy equation of state $w$, and amplitude of matter fluctuations $\sigma_8$. Interestingly, our analysis shows that neglecting lensing convergence might contribute to exacerbate the discrepancies between low and high redshift probes concerning $H_0$ and $\sigma_8$. We carried out our investigation by using different configurations for the galaxy distribution and we observed no significant change in our conclusions.

\section*{Acknowledgements}

We are grateful to Jose Beltr\'an Jim\'enez and J. Bayron Orjuela-Quintana for carefully reading the manuscript and providing comments. DF acknowledges support from the programme {\it Ayudas para Financiar la Contrataci\'on Predoctoral de Personal Investigador (ORDEN EDU/601/2020)} funded by Junta de Castilla y Le\'on and European Social Fund. DF acknowledge support from the {\it Atracci\'on del Talento Cient\'ifico en Salamanca} programme, from Project PGC2018-096038-B-I00 funded by the Spanish "Ministerio de Ciencia e Innovación" and FEDER “A way of making Europe", and {\it Ayudas del Programa XIII} by USAL. WC acknowledges financial support from the S\~{a}o Paulo Research Foundation (FAPESP) through grant \#2021/10290-2. This research was supported by resources supplied by the Center for Scientific Computing (NCC/GridUNESP) of the S\~{a}o Paulo State University (UNESP).   

\bibliographystyle{JHEPmodplain}
\bibliography{biblio2}


\end{document}